\documentclass{aa}

\usepackage{graphicx}
\usepackage{txfonts}
\usepackage{hyperref}

\usepackage{bookmark}

\usepackage{graphicx}	
\usepackage{amsmath}	

\usepackage{longtable}
\usepackage{booktabs}
\usepackage{xcolor}
\usepackage{xspace}
\usepackage{multirow}

\defcitealias{Wen2024ApJS..272...39W}{WH24}

\begin{document} 

\title{A catalog to unite them all: REGALADE, a revised galaxy compilation for the advanced detector era}
\author{Hugo Tranin\inst{1,2}, Nadejda Blagorodnova\inst{1,2,3}, Marco A. Gómez-Muñoz\inst{1,2},  Maxime Wavasseur\inst{1,2}, Paul J. Groot\inst{4,5,6,7}, Lloyd Landsberg\inst{5}, Fiorenzo Stoppa\inst{8}, Steven Bloemen\inst{4}, Paul M. Vreeswijk\inst{4}, Dani\"elle L.A. Pieterse\inst{4}, Jan van Roestel\inst{9}, Simone Scaringi\inst{10,11}, Sara Faris\inst{12}
}

\institute{Institut de Ciències del Cosmos (ICCUB), Universitat de Barcelona (UB), c. Martí i Franquès, 1, 08028, Barcelona, Spain
\\e-mail: \texttt{htranin@icc.ub.edu}
\and
Departament de Física Quàntica i Astrofísica (FQA), Universitat de Barcelona (UB), c. Martí i Franquès, 1, 08028, Barcelona, Spain
\and
Institut d'Estudis Espacials de Catalunya (IEEC), Edifici RDIT, Campus UPC, 08860 Castelldefels (Barcelona), Spain
\and
Department of Astrophysics/IMAPP, Radboud University, PO Box 9010, 6500 GL Nijmegen, The Netherlands
\and
Department of Astronomy, University of Cape Town, Private Bag
X3, Rondebosch, 7701, South Africa
\and
South African Astronomical Observatory, P.O. Box 9, Observatory,
7935, South Africa
\and
The Inter-University Institute for Data Intensive Astronomy, University of Cape Town, Private Bag X3, Rondebosch, 7701, South Africa
\and
Astrophysics sub-Department, Department of Physics, University of Oxford, Denys Wilkinson Building, Keble Road, Oxford, OX1 3RH, UK
\and
Anton Pannekoek Institute for Astronomy, University of Amsterdam, P.O. Box 94249, 1090 GE Amsterdam, The Netherlands
\and
Department of Physics, Centre for Extragalactic Astronomy, Durham University, South Road, Durham DH1 3LE, UK
\and
INAF – Osservatorio Astronomico di Capodimonte, Salita Moiariello 16, I-80131 Naples, Italy
\and
The School of Physics and Astronomy, Tel Aviv University, Tel Aviv 69978, Israel
}

\abstract
    {Many applications in transient science, gravitational wave follow-up, and galaxy population studies require all-sky galaxy catalogs with reliable distances, extents, and stellar masses. However, existing catalogs often lack completeness beyond $\sim 100$ Mpc, suffer from stellar contamination, or do not provide homogeneous stellar mass estimates and size information.}
    {Our goal is to build a high-purity, high-completeness, all-sky galaxy catalog out to 2,000 Mpc, specifically designed to support time-domain and multi-messenger astrophysics.}
    {We combined major galaxy catalogs and deep imaging surveys -- including the Legacy Surveys, Pan-STARRS, DELVE, and SDSS -- and added spectroscopic, photometric, and redshift-independent distances. We cleaned the sample using the Gaia catalog to remove stars and visually inspected all ambiguous cases below 100\,Mpc through a classification platform that gathered 27,000 expert votes. Stellar masses were estimated using optical and mid-infrared profile-fit photometry, and we improved the accuracy of photometric distances by combining multiple independent estimates.}
    {The resulting catalog, REGALADE, includes nearly 80 million galaxies with distances under 2,000\,Mpc. It provides stellar masses for 88\% of the sample and ellipse fits for 80\%. REGALADE is more than 90\% complete for galaxies contributing 50\% of the total $r$-band luminosity out to 360\,Mpc. In science tests, it recovers 60\% more known supernova hosts, doubles the number of low-luminosity transient hosts, and identifies more reliable hosts for ultraluminous and hyper-luminous X-ray sources.}
    {REGALADE is one of the most complete and reliable all-sky galaxy catalog to date for the nearby Universe, built for real-world applications in transient and multi-messenger astrophysics. The full dataset, visual classifications, and code will be released to support broad community use.}
\keywords{Galaxies: general -- Galaxies: distances and redshifts -- Catalogs -- Surveys -- Supernovae: general -- Gravitational Waves}
\titlerunning{REGALADE: making the most of existing galaxy catalogs}
\authorrunning{H. Tranin et al.}

\maketitle
%

\section{Introduction}\label{sec:introduction}

A complete census of nearby galaxies is essential for a broad range of astrophysical studies, including many time-domain and multi-messenger applications. In gravitational wave (GW) follow-ups, for example, galaxy-targeted strategies rely on volume-limited samples to prioritize likely host candidates. With thousands of supernovae discovered each year by surveys such as ZTF \citep{Bellm2019}, ATLAS \citep{Tonry2018PASP..130f4505T}, GOTO \citep{goto2020MNRAS.497..726G}, and BlackGEM \citep{Groot}, and with the upcoming Vera C. Rubin Observatory's Legacy Survey of Space and Time (LSST; \citealt{LSST2019}) expected to produce up to 10 million alerts per night, rapid classification of transients is becoming a major challenge. Host galaxy distances provide the key information needed to estimate absolute magnitudes and classify transients reliably and in real time \citep{Muthukrishna2019, Moller2020}.

Early galaxy catalogs such as the Third Reference Catalogue of Bright Galaxies (\citealt{deVaucouleurs1991rc3..book.....D}) provided uniform photometry, sizes, and redshifts for bright galaxies with large angular sizes (typically $>1$ arcmin) but were limited in depth and completeness. Several large redshift surveys have significantly addressed these limitations. The Sloan Digital Sky Survey (SDSS; \citealt{York2000AJ....120.1579Y, Abdurrouf2022}) has provided high-quality photometry and spectroscopy for millions of galaxies; however, it only covers $\sim$35\% of the sky, in the northern hemisphere. The Dark Energy Survey (DES; \citealt{Abbott2021ApJS..255...20A}) and more recently the Legacy Surveys \citep{Dey2019AJ....157..168D} extend photometric coverage to deeper magnitudes and larger regions but likewise do not achieve full-sky coverage. As a result, important sky regions -- especially at low Galactic latitudes and in the southern hemisphere -- remain underrepresented in galaxy compilations.

 Efforts to build complete compilations -- such as HyperLEDA \citep{Makarov2014A&A...570A..13M} and the NASA Extragalactic Database (NED) Local Volume Sample (NED-LVS; \citealt{Cook2023ApJS..268...14C}) -- have merged spectroscopic, photometric, and redshift-independent distances across multiple catalogs. Initially focused on nearby ($\lesssim 80$ Mpc) galaxies with optical and 21cm data, HyperLEDA has expanded to include over 3 million galaxies with parameters such as morphology, inclination, rotational velocity, magnitudes, and redshifts \citep{Makarov2014A&A...570A..13M, Kovlakas2021MNRAS.506.1896K}, while NED-LVS contains 2 million galaxies with distances out to 1000\,Mpc including multiple distance estimates, multiwavelength photometry, stellar mass, and star formation rate estimates. These resources are widely used to find the host of transient or GW events, where volumetric completeness is critical.

 Among recent efforts to support multi-messenger astronomy, the GLADE (Galaxy List for the Advanced Detector Era; \citealt{Dalya2018MNRAS.479.2374D}) and HECATE (Heterogeneous Catalog of Nearby Galaxies; \citealt{Kovlakas2021MNRAS.506.1896K}) catalogs have been widely adopted for host identification and statistical analyses within a distance $D < 200$\,Mpc. GLADE and its successor GLADE+ \citep{Dalya2022MNRAS.514.1403D} compile galaxies from multiple surveys and are complete in $B$-band luminosity of massive galaxies (giving 90\% of the total K-band luminosity) up to $\gtrsim$100\,Mpc, with extended coverage to 800\,Mpc. HECATE is a value-added catalog focusing on galaxies within 200\,Mpc, with homogeneously reprocessed distances, star formation rates, and stellar masses, including multiwavelength data from optical to far infrared.

 Despite these efforts, significant incompleteness persists beyond $z \sim 0.1$ as well as for dwarf galaxies or low surface brightness systems. Furthermore, few catalogs include galaxy extent data, despite the fact that such data are crucial for associating and studying populations of extragalactic transients, often occurring in the galaxies' outskirts. Representing each galaxy as an ellipse, based on its apparent major axis, axis ratio, and orientation, allows for more accurate probabilistic host association. This gap limits our ability to compute unbiased volumetric rates or to prioritize follow-up targets efficiently.

 Here we present the Revised Galaxy List for the Advanced Detector Era (REGALADE), a new all-sky catalog of nearly 80 million galaxies out to 2,000\,Mpc. REGALADE combines redshift and distance data from curated sources with homogenized multiband photometry, extent parameters, and systematic removal of stellar and spurious contaminants. We validate the catalog across multiple science applications -- including transient host recovery, studies of ultra- and hyper-luminous X-ray sources (ULX and HLX), and GW follow-up -- and benchmark it against existing resources such as GLADE+ and HECATE.

 The choice of a 2000\,Mpc ($z \lesssim 0.37$) threshold is motivated by the sensitivity range of current and upcoming facilities, as well as the intrinsic luminosities of key transient classes. For GW detections, this distance encompasses the design sensitivity horizon of Advanced LIGO and Virgo for neutron star–black hole mergers \citep{Abbott2020LRR....23....3A}, that is 200–300\,Mpc for binary neutron star mergers and beyond for neutron star–black hole systems. Third-generation detectors (e.g., Cosmic Explorer; \citealt{CosmicExplorer}, Einstein Telescope; \citealt{Einstein}, LISA; \citealt{LISA2023}) will extend sensitivity to several gigaparsecs. From the electromagnetic side, this range aligns with the single-visit depth of LSST ($r \sim 24.5$; \citealt{LSST2009arXiv0912.0201L}), which is sufficient to detect kilonovae such as AT2017gfo out to $z \sim 0.1$–0.2. Long gamma-ray bursts with optical afterglows have also been observed in this redshift range (e.g., GRB 130427A at $z = 0.34$; \citealt{Perley2014ApJ...781...37P}). Tidal disruption events, including those identified by ZTF and Pan-STARRS, have been detected out to $z \sim 0.2$–0.3 \citep{vanvelzen2020SSRv..216..124V, bricman2023}, and super-luminous supernovae can be detected out to $z \gtrsim 1$ (e.g., \citealt{Quimby2011Natur.474..487Q}). Advancing our understanding of the lesser-known classes of intermediate luminosity optical transients -- such as luminous red novae \citep{Pastorello2019b, Blagorodnova2021} -- also depends on having a comprehensive galaxy catalog extending to at least 300\,Mpc, to ensure a rapid and complete identification of these events \citep{Karambelkar2023}. A complete galaxy catalog out to 2,000\,Mpc is thus crucial for enabling targeted follow-up, assigning hosts, and estimating intrinsic rates of the transient zoo across the full observable volume of present and upcoming surveys.

\begin{table*}
\caption{\label{tab:ranking}Main properties of catalogs used in this study, including their crossmatch order, coverage, ellipse, and distance ranking.}
\centering
\resizebox{2\columnwidth}{!}{
\begin{tabular}{l l r c c r r r r r r l}
\hline
Index & Catalog & N & Ellipse & Primary & Sky cov. & $|b|$ ($^\circ$) & Dec ($^\circ$) & Parameter space & $E_\mathrm{rank}$ & $D_\mathrm{rank}$ & Reference\\
\hline
0 & SGA & 383,343 & Yes & No & 0.50 & $>20$ & $>-65$ & $R_1>10$\,arcsec & 1 & 6 &  \citet{Moustakas2023}\\
1 & GLADE1 & 1,914,351 & Yes & No & 0.99 & - & - & - & 3 & 10 & \citet{Dalya2018MNRAS.479.2374D}\\
2 & HECATE & 204,487 & Yes & No & 0.92 & - & - & $D<200$\,Mpc & 4 & 7 & \citet{Kovlakas2021MNRAS.506.1896K}\\
3 & DESI PV & 4,440,551 & Yes & No & 0.43 & $>20$ & $>-30$ & $D<700$\,Mpc & 5 & 8 & \citet{Saulder2023MNRAS.525.1106S}\\
4 & DESI DR1 & 13,646,368 & Yes & Yes & 0.37 & $>20$ & $>-20$ & $r<20.3$\,mag & 6 & 1 & \citet{DESI2025}\\
5 & Cosmicflows & 55,868 & No & Yes & 0.64 & - & - & $D<500$\,Mpc & - & 2 & \citet{Tully2023ApJ...944...94T}\\
6 & NED-LVS-D & 12,250 & No & Yes & 0.43 & - & - &  $D<1000$\,Mpc & - & 3 &\citet{Cook2023ApJS..268...14C}\\
7 & NED-LVS-$z_{sp}$ & 1,352,895 & No & Yes & 0.95 & - & - & $D<1000$\,Mpc & - & 4 &\citet{Cook2023ApJS..268...14C}\\
8 & NED-LVS-rest & 688,423 & No & No & 0.98 & - & - & $D<1000$\,Mpc & - & 5 &\citet{Cook2023ApJS..268...14C}\\
9 & GLADE+ & 22,754,931 & No & No & 0.97 & - & - & - & - & 9 &\citet{Dalya2022MNRAS.514.1403D}\\
10 & LS DR9 & 110,142,409 & Yes & No & 0.5  & $>20$ & $>-65$ & $r<22.0$\,mag & 7 & 13  & \citet{Zou2022RAA....22f5001Z}\\
11 & Pan-STARRS & 40,440,659 & Yes & No & 0.75 & - & $>-30$ & $r<21.0$\,mag & 8 & 11 & \citet{Chambers2016}\\
12 & SDSS & 7,989,000 & Yes & No & 0.45 & - & $>-25$ & $r<21.0$\,mag & 9 & 16 & \citet{Abdurrouf2022}\\
13 & GSC blue & 97,696,980 & Yes & No & 0.99 & - & - & $r<22.0$\,mag & 10 & 12  & \citet{Lasker2008AJ....136..735L}\\
14 & LS DR10 & 55,595,475 & No & No & 0.63 & $>20$ & $<+30$ & $r<22.0$\,mag & - & 14 & \citet{Wen2024ApJS..272...39W}\\
15 & DELVE & 53,665,084 & Yes & No & 0.51 &  $>10$ & $<+30$ & $r<21.0$\,mag & 2 & 15 & \citet{Drlica2022ApJS..261...38D}\\
\hline
\end{tabular}
}
\tablefoot{Number counts refer to input catalog size after duplicate removal (Section \ref{sec:duplicate}). "Ellipse" and "Primary" indicate the presence of ellipse parameters and primary distance indicators. "Sky cov." is the fractional occupation of a HEALPix map (Nside=32). "$|b|$" and "Dec" precise sky coverage by giving the range of absolute Galactic latitude and declination, respectively. "Parameter space" gives the volume, size or flux limit of each survey. $E_\mathrm{rank}$ and $D_\mathrm{rank}$ indicate the priority order for adopting the best input ellipse and distance, respectively.}
\end{table*}

\section{Input catalogs}
\label{sec:dataset}

\subsection{Curated galaxy catalogs}

To begin our compilation, we collected the most widely used galaxy catalogs in transient and extragalactic studies. These include HECATE \citep{Kovlakas2021MNRAS.506.1896K}, the Siena Galaxy Atlas (SGA, \citealt{Moustakas2023}), GLADE, and GLADE+ \citep{Dalya2018MNRAS.479.2374D, Dalya2022MNRAS.514.1403D}, as well as the updated NASA/IPAC Extragalactic Database (NED) Local Volume Sample (NED-LVS; \citealt{Cook2023ApJS..268...14C, NEDLVS2025}). We also included CosmicFlows \citep{Tully2016AJ....152...50T, Tully2023ApJ...944...94T, Valade2024NatAs...8.1610V}, providing redshift-independent distances to complement redshift-based estimates, especially in the local Universe where peculiar velocities dominate. To extend coverage, we added galaxy samples from deep imaging surveys using custom queries (Sections \ref{sec:deep1}, \ref{sec:deep2}, \ref{sec:ps1}). A summary of all input catalogs is given in Table \ref{tab:ranking}. Throughout this work, redshift-based distances assume a flat Lambda cold dark matter cosmology with $H_0 = 71$ km/s/Mpc, $\Omega_m = 0.27$, and $\Omega_\Lambda = 0.73$ \citep{Spergel2003ApJS..148..175S}.

The original GLADE catalog (hereafter GLADE1\footnote{\url{https://vizier.cds.unistra.fr/viz-bin/VizieR?-source=VII/275}}; \citealt{Dalya2018MNRAS.479.2374D}) is an all-sky compilation optimized for GW counterpart searches. It merges HyperLEDA, the 2MASS extended source catalog (XSC, \citealt{Skrutskie2006AJ....131.1163S}), and GWGC \citep{White2011CQGra..28h5016W}, to achieve higher completeness within $D < 100$\,Mpc; it includes galaxy extent for nearly all of its 1.9 million entries. The updated GLADE+ adds deeper and more recent surveys to improve completeness at higher redshift but lacks extent information, which limits its usefulness for transient host identification.

NED-LVS \citep{Cook2023ApJS..268...14C}) is a curated compilation offering homogenized redshift and distance estimates. From the January 2025 release, containing 2.06 million objects (including 1.98 million galaxies), we defined the following three subsamples, to track distance reliability across the catalog:

\begin{itemize}

\item NED-LVS-D: galaxies with reliable redshift-independent distances. They were selected for having a primary distance indicator (\texttt{ziDist\_indicator} = "P") or a robust aggregated value (\texttt{zidist\_method} = "Med" or "Wavg");
\item NED-LVS-$z_\mathrm{sp}$: galaxies only with spectroscopic redshifts;
\item NED-LVS-rest: galaxies in the remaining sample.
\end{itemize}

\subsection{Legacy Surveys and DESI catalogs}
\label{sec:deep1}

Among the largest deep surveys available are SDSS \citep{Abdurrouf2022}, Pan-STARRS \citep{Chambers2016}, and the Legacy Surveys \citep{Dey2019AJ....157..168D}, each covering complementary regions of the sky (Table~\ref{tab:ranking}). We included galaxy catalogs derived from the Legacy Surveys DR9 (LS DR9; \citealt{Dey2019AJ....157..168D}) and DR10 (LS DR10), using the extensive public datasets by \citet{Zou2022RAA....22f5001Z} and \citet{Wen2024ApJS..272...39W} (hereafter \citetalias{Wen2024ApJS..272...39W}). The LS DR9 comprises three major projects: the Dark Energy Camera Legacy Survey (DECaLS), the Beijing-Arizona Sky Survey (BASS), and the Mayall z-band Legacy Survey (MzLS). DR10 extends these data by including observations from newly covered regions with DECaLS, as well as new i-band photometry in some areas previously included in LS DR9. Our selection included all of their galaxies at $z_{\mathrm{ph}} < 0.4$.

Other catalogs based on Legacy Surveys but related to the observations of the Dark Energy Spectroscopic Instrument (DESI; \citealt{Levi2019}) included the target selection for the DESI peculiar velocity survey (DESI-PV; \citealt{Saulder2023MNRAS.525.1106S}) and DESI DR1 \citep{DESI2025}. The latter provides over 18 million spectroscopic redshifts of galaxies, making it the largest spectroscopic galaxy sample released to date. We selected galaxies following these criteria ensuring reliable redshift and classification: \texttt{ZCAT\_PRIMARY} \& \texttt{ZWARN} = 0 \& $\texttt{Z}>0.002$ \& $\texttt{SPECTYPE}=\texttt{"GALAXY"}$\footnote{\texttt{ZCAT\_PRIMARY} flags the unique entry corresponding to the canonical redshift measurement for each target, \texttt{ZWARN = 0} ensures that no issues were detected during redshift fitting, and \texttt{SPECTYPE} specifies the object classification (star, galaxy, or quasar) derived from spectral template fitting (Section 3.1.2. of \citealt{DESI2025}).}. Furthermore, after visual inspection of very low redshift objects, we noticed a parameter space with a majority of likely stars; therefore, we discarded $\sim$ 400,000 objects with $\texttt{Z} < 0.01/\max(\texttt{SHAPE\_R}, 0.02)$. The final sample contains 13.7 million galaxies.

\subsection{Deep survey galaxy tables}
\label{sec:deep2}

To extend coverage beyond the Legacy Surveys footprint, we extracted galaxies directly from SDSS DR17 \citep{Abdurrouf2022} and the Dark Energy Camera Local Volume Exploration Surveys (DELVE; \citealt{Drlica2022ApJS..261...38D}) -- overlapping with the Legacy Surveys via DECaLS -- using custom queries (Section \ref{sec:sql}). In SDSS, we selected a de Vaucouleurs size-limited and flux-limited sample\footnote{Specifically, based on the de Vaucouleurs morphological fit, we required an $r$-band model radius of $\texttt{deVRad\_r} > 2$ arcsec and a $g$-band model magnitude of $\texttt{deVMag\_g} < 21$ mag (\url{https://www.sdss4.org/dr14/algorithms/classify/}).} of $\sim 8$ million galaxies flagged as "clean" (Appendix \ref{sec:sql}). We retrieved de Vaucouleurs profile fits and photometric redshifts for all galaxies. Visual inspection revealed no significant improvement in ellipse fitting when using the exponential profile. We excluded objects whose fitted size was inconsistent with their apparent magnitude ($\log(\texttt{deVRad\_r}) > (25 - \texttt{deVMag\_r})/5$).

DELVE extends deep DECaLS photometry to regions previously little covered. Our selection was based on photometric and morphological criteria, targeting sources with $r$-band magnitude brighter than 21 and classified as extended. We added DELVE photometric redshifts when available (i.e., only for galaxies having clean photometry in all $g,r,i,z$ bands, \citealt{Drlica2022ApJS..261...38D}).

\subsection{Custom queries to imaging surveys}
\label{sec:ps1}

We also extracted galaxies from Pan-STARRS and the Guide Star Catalog (GSC 2.4.2; \citealt{Lasker2008AJ....136..735L}) to improve coverage near the Galactic plane. From Pan-STARRS DR1, we selected extended sources using \( N_s > 0 \), \( i_K < 21 \), \( z - z_K > 0.3 \) and \( g - g_K > 0.3 \), where $N_s$ is the number of stack detections, and $(g,r,i,z)$ and ($g_K,r_K,i_K,z_K$) stand for PSF and adaptive-aperture Kron magnitudes, respectively \citep{Chambers2016}. Photometric redshifts were taken from the dedicated catalog of \citet{Tarrio2020A&A...642A.102T}. Outside the Galactic plane ($|b| > 20^\circ$), Pan-STARRS galaxies often have Sersic or de Vaucouleurs profile measurements from the \texttt{ForcedGalaxyShape} table \citep{Flewelling2020}. Closer to the plane, however, many selected objects were stars or diffraction-spike artifacts near bright stars. These spurious sources, which are absent from the photometric redshift catalog, tend to cluster near bright stars. Using the \textit{internal match} function of TOPCAT, we grouped detections without redshift within a 30'' radius to identify clusters containing at least three members, and excluded all objects located within 1' of them.

To verify that this filtering step effectively removes spurious detections, we visually inspected 50 randomly selected sources with $z<0.2$ that remain after filtering in the region $|b| < 10^\circ$. The large majority appear to be genuine galaxies, with fewer than ten stars or ambiguous detections. Conversely, among 700 randomly inspected objects removed by this filter (out of $\sim$2 million in total), four were identified as genuine galaxies, corresponding to a fraction of less than 1.5\% of removed objects (95\% confidence).

For GSC, originally developed for astrometry but including millions of extended sources with ellipse parameters, we selected nearby blue galaxies using a custom color selection (Section \ref{sec:sql}). This was aimed at recovering star-forming dwarf galaxies, which are likely hosts of super-luminous supernovae \citep{Perleu2016ApJ...830...13P} but may be underrepresented in other surveys. All objects from these custom-extracted samples were crossmatched with photometric redshift catalogs from \citet{Tarrio2020A&A...642A.102T} and \citet{Duncan2022MNRAS.512.3662D}. We retained sources with $z_\mathrm{ph} < 1$ for GSC and $z_\mathrm{ph} < 0.5$ for Pan-STARRS. We also excluded sources having $z_\mathrm{ph} < 10^{(r_K - r)/2.5 - 1}$, confirmed as stars by visual inspection.

\section{Methods}
\label{sec:methods}

\subsection{Size and distance calibration}
\label{sec:calib}

Among all input catalogs, ten have some information about the extent of their objects (Table \ref{tab:ranking})\footnote{Different catalogs define galaxy extent using various criteria, typically either a surface-brightness threshold (in mag arcsec$^{-2}$) or a fraction of the total light profile. In REGALADE, sizes from these catalogs were recalibrated to a common system, adopting as reference the SGA definition: the $r$-band isophotal radius at $\mu = 26~\mathrm{mag~arcsec}^{-2}$.}. For self-consistency, we systematically crossmatched these catalogs against each other to recompute their ellipse $E$ parametrized by semimajor axis $R_1$, semiminor axis $R_2$ and position angle $PA$ measured positively from north to east. A comparison between different catalog ellipses and distance estimates is further illustrated in Appendix \ref{sec:ap1}. Table \ref{tab:calib} lists input parameter names and equations used to obtain the final ellipse. For example, \texttt{D26} from SGA is the major axis in arcmin, so the semimajor axis in arc seconds is simply $R_1 = 60~$\texttt{D26}/2.

The comparison with DESI spectroscopic distances showed better agreement for the smaller catalogs, which are tailored to accurate measurements of large, bright galaxies within $D < 1000$ Mpc. We therefore defined a catalog as a primary distance reference if its distances are derived from redshift-independent methods or spectroscopic redshifts (catalogs 4, 5, 6, 7 in Table \ref{tab:ranking}). Larger catalogs using photometric redshifts, such as Pan-STARRS, GLADE+, or SDSS, also showed reasonable consistency, albeit with more scatter.

\subsection{Duplicate removal}
\label{sec:duplicate}
Once all input catalogs had their $R_1$, $R_2$, and $PA$ parameters homogenized, they underwent a series of preprocessing steps. First, galaxies lacking ellipse parameters or with $R_1$ values smaller than 3 arc seconds were reset so that both $R_1$ and $R_2$ equal 3 arc seconds. This threshold corresponds to the typical apparent size of a galaxy at the catalog's distance limit. Specifically, the angular size is given by \(\theta \approx r_\mathrm{kpc}/D_\mathrm{ang} = 1.4 \times 10^{-5}\,\mathrm{radians} \approx 3\,\mathrm{arcsec}\), assuming a Milky Way-like physical radius of \(r_\mathrm{kpc} = 15\,\mathrm{kpc}\) and an angular diameter distance \(D_\mathrm{ang} = 1060\,\mathrm{Mpc}\). This distance corresponds to a luminosity distance of $D=2000$\,Mpc in the adopted cosmology. Next, duplicate entries were identified and removed through an internal sky matching process. This matching uses a variable radius defined as the maximum between the galaxy's radius toward the matched object and 3 arc seconds (Section \ref{sec:xmatch}). 
When duplicates were found, preference was given to entries with available distance measurements and larger $R_1$ values.

\subsection{Crossmatch between catalogs}
\label{sec:xmatch}

 The positional uncertainty of a galaxy is closely linked to its apparent size. Large galaxies at $D<100$ Mpc -- particularly irregular ones -- often have poorly defined centers, leading to shifts in their recorded positions of more than 10 arc seconds between different catalogs. In contrast, distant galaxies generally appear smaller, and their positions can be measured with sub-arc-second precision in deep surveys. To account for this variable positional uncertainty, we performed elliptical matching between each pair of input catalogs using the \texttt{match\_to\_catalog\_sky} function from \texttt{astropy}. Two galaxies were considered a match if they were closer than 5 arcsec or if their ellipses each encompassed the center of the other. For two ellipses \(a\) and \(b\), this matching criterion of the angular separation $s$ is expressed as

 $$s < \max\left(5, \min_{a,b} \left( \frac{R_1 R_2}{\sqrt{R_1^2 \sin^2(\phi) + R_2^2 \cos^2(\phi)}} \right)\right),$$

 where \(\phi = \theta - PA\), with \(\theta\) denoting the oriented angle between the central coordinates of \(a\) and \(b\), and \(PA\) representing the position angle of the ellipse.

 The matching algorithm proceeded by sequentially crossmatching input catalogs in the order shown in Table \ref{tab:ranking}. Matched entries were merged into a single record and unmatched entries were added to the compilation. For each record, we calculated the one-point trimmed mean of the available distances ($D_\text{tmean}$)\footnote{Average computed after removing the smallest and largest values (i.e., trimming one point from each end of the sorted sample).}. The trimmed mean offers a computationally efficient estimate and effectively corresponds to the median when fewer than five distances are available. Additionally, we provided the best input distance $D_\text{input}$ in a separate column, selected according to a strict priority order among the catalogs as defined in Section \ref{sec:calib}. We defined

 $$D = \begin{cases}
 D_\text{input} & \text{if }N_D<3\text{ or primary distance reference}\\
 D_\text{tmean} & \text{otherwise,}
 \end{cases}$$
 
  where $N_D$ is the number of available distances. Only entries with $D<2000$\,Mpc were retained. 

 Similarly, the best ellipse for each galaxy was retained based on a distinct priority ranking. The resulting compilation contained 88 million entries.

\subsection{Photometry retrieval}
\label{sec:photo}

We retrieved photometry for our compilation in three stages.
First, in the optical bands we adopted Kron magnitudes from Pan-STARRS ($g_K,r_K,i_K,z_K$) and, when available, added measurements from the catalogs of \citet{Zou2022RAA....22f5001Z} and \citetalias{Wen2024ApJS..272...39W}, both based on Legacy Surveys DR9 and DR10. These two sources provided $g,r,i,z,W1,W2$ photometry along with quality flags and morphological parameters. Photometry from Zou et al. and \citetalias{Wen2024ApJS..272...39W} is already corrected for Galactic extinction using the \texttt{mw\_transmission} values from the Legacy Surveys, based on the dust maps of \citet{Schlegel1998ApJ...500..525S}. In contrast, Pan-STARRS and DELVE data are not extinction-corrected. For those, we interpolated $E(B-V)$ from the \citet{Schlegel1998ApJ...500..525S} maps and applied extinction coefficients from \citet{Yuan2013MNRAS.430.2188Y} to correct $g,r,i,z,W1,W2$ magnitudes.

Second, we filled in mid-infrared coverage for galaxies not present in the Legacy Surveys by crossmatching with the AllWISE catalog \citep{Cutri2012yCat.2311....0C}, retrieving $K$, $W1$, and $W2$ magnitudes (converted to AB), and with the 2MASS XSC to obtain the isophotal $K\_\mathrm{ext}$ magnitude. To convert WISE point-source photometry into profile-fit magnitudes, we applied proxy relations described in Section \ref{sec:w1ext_proxies}. 

As a third step, we crossmatched the full sample with Gaia DR3 to retrieve $G$ and $BP$ magnitudes, angular separations, and proper motions. These quantities were used to identify stellar contaminants (Section \ref{sec:contam}).

\subsection{Removal of contaminants}
\label{sec:contam}

Galaxy catalogs often include contaminants such as stars, image artifacts, HII regions, and Galactic gas. To reduce such contamination, we applied a multistep filtering process based on catalog multiplicity, position, photometry, and morphology.

We first flagged “orphans” -- sources present in only one input catalog -- and removed those overlapping larger galaxies (diameter $>$40''), which are often HII regions or substructures. Orphans near the Galactic plane were also excluded using a latitude-dependent mask defined as $|b|<15\cos^2(l/2)+1$, since they are likely stars or artifacts in crowded stellar fields. This empirical mask, originally derived from visual inspection of multiple fields, closely follows a density contour that encloses 95\% of Milky Way stars in the thin and thick disks, as modeled by two exponential components with parameters from \citet{Juric2008}.

\begin{figure}
    \centering
    
    \includegraphics[width=\linewidth]{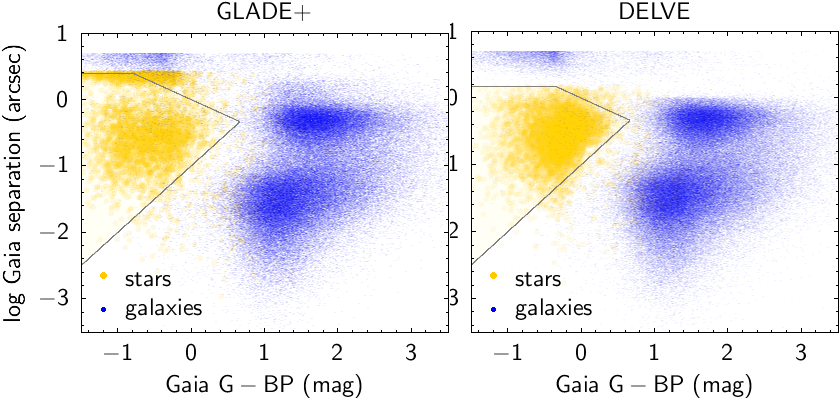}
    \includegraphics[width=0.7\linewidth]{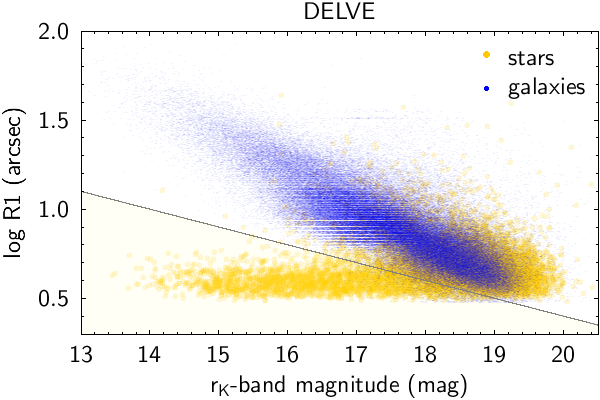}
    \caption{Parameter space used to identify stellar contaminants, for GLADE+ (top left) and DELVE (top right, bottom) objects. Galaxies (blue dots) shown in these plots are objects that appear in at least four input catalogs. Stars (yellow dots) are identified as objects in at most two catalogs that have a Gaia counterpart located within a separation of \( s < 2 \) arc seconds and exhibit a proper motion greater than 5 mas/yr. The yellow-shaded region highlights the parameter space selected to isolate Gaia stars, which for the top panels is defined by the condition
\(
G - BP - 1 < \log\left(s/\text{arcsec}\right) < \min\left(\log(s_\text{max}),~ -0.5(G - BP)\right),
\)
where \( s_\text{max} = 2.5 \) for GLADE+ and 1.5 for the other catalogs. For the bottom panel, the selection follows the condition
\(
\log\left(R_1/\text{arcsec}\right) < (24 - r_K)/10
\).}
    \label{fig:angdist_color}
\end{figure}

To identify stellar contaminants, we used the Gaia source separation and $G - BP$ color, which cleanly separates stars and quasars from galaxies \citep{Delchambre2023}. Sources within a defined region in this parameter space (Figure \ref{fig:angdist_color}, top) were excluded, as were objects whose angular sizes were too small for their $r_K$ magnitude (Figure \ref{fig:angdist_color}, bottom). However, we retained compact sources if they appeared in at least four input catalogs or if they matched a Legacy Survey object from \citet{Duncan2022MNRAS.512.3662D} that was classified as a galaxy (i.e., with $p_{\rm star} < 0.5$). In other words, compact morphology alone did not trigger removal when the object had an external galaxy classification. These criteria removed 6.7 million stellar contaminants.

We also discarded orphans undetected in all optical and infrared bands, unless they were present in HECATE or Cosmicflows, which include low-surface-brightness and Local Group galaxies. Lastly, we removed 1,078 DESI-PV orphan sources that did not match any other galaxy ellipse but formed clusters of more than ten objects within 2' of each other\footnote{These groups were again identified using the \textit{internal match} function of TOPCAT. Visual inspection of 50 removed sources from different groups revealed only spurious detections near bright sources.}. Despite the cleaning filters applied, a small fraction of spurious sources (primarily stars or artifacts) may still remain in REGALADE. These contaminants are most likely among objects detected in only a subset of the large all-sky surveys and absent from smaller, targeted compilations of nearby galaxies. To assist users in identifying such potentially unreliable entries, we include a flag, \texttt{f\_reliability}, set to 1 for galaxies detected in exclusively one or two lower-reliability catalogs (specifically catalogs 3, 8, 10, 11, 13, 14 and 15 in Table \ref{tab:ranking}). Sources with $\texttt{f\_reliability} = 1$ (27\% of REGALADE) should therefore be treated with caution in applications requiring a high level of purity.

\begin{figure}
    \centering
    \caption{Comparison of galaxies in input catalogs (left) and in the final REGALADE compilation (right), for a 2.6' field centered at RA = 180.19$^\circ$, Dec = $-$0.03$^\circ$. The background is a ($g,r,i,z$) composite from Legacy Surveys DR10.}
    \includegraphics[width=\linewidth]{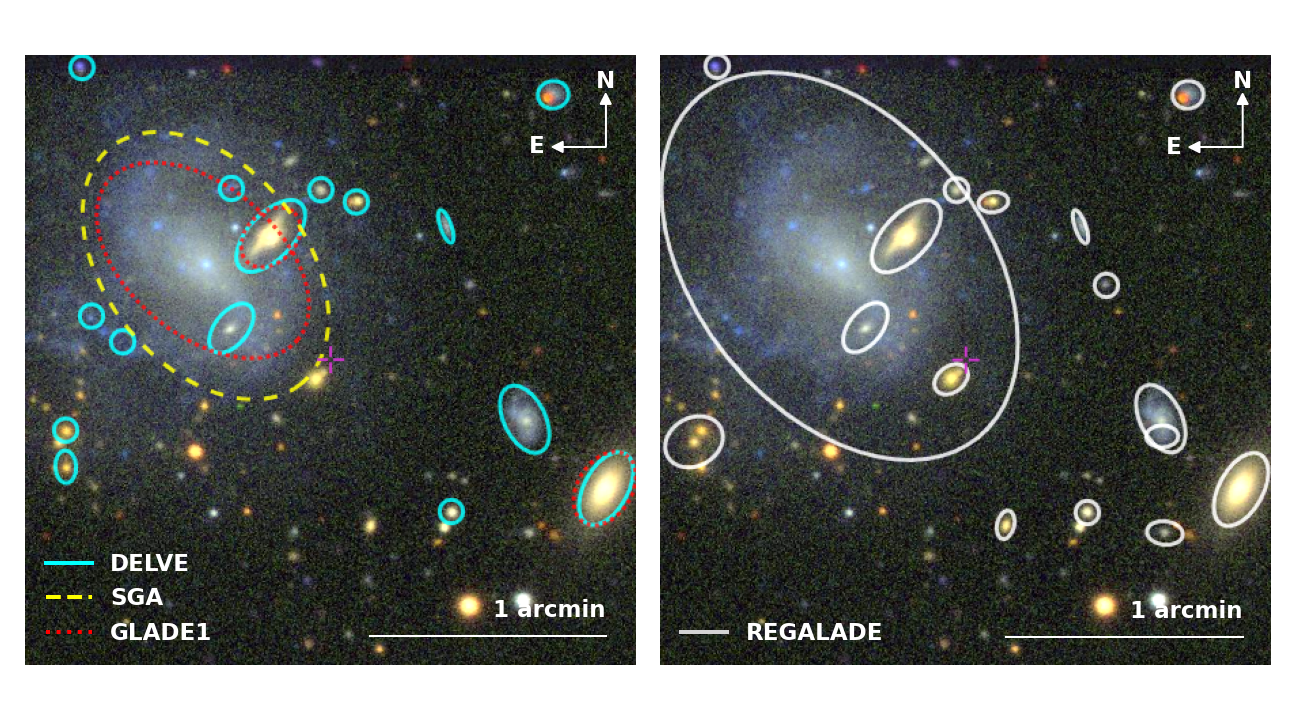}
    \label{fig:skyplot}
\end{figure}

Figure \ref{fig:skyplot} shows the impact of these filters in a 2.6' field at RA = 180.19$^\circ$, Dec = $-$0.03$^\circ$, including the large galaxy UGC 6998. The resulting sample is significantly cleaner and includes consistent redshifts, galaxy profiles, and ellipse measurements from DELVE, GLADE1, and SGA, thanks to the combination of all input catalogs.

\subsection{Stellar mass estimates}
\label{sec:mass_method}

We compared stellar mass estimates from several sources (e.g., \citealt{Parkash2018ApJ...864...40P, Kovlakas2021MNRAS.506.1896K, Zou2022RAA....22f5001Z, Blanton2007AJ....133..734B, Wen2024ApJS..272...39W}), finding \citetalias{Wen2024ApJS..272...39W} to offer the best tradeoff between accuracy (0.15 dex scatter compared to COSMOS2015 values, \citealt{Wen2024ApJS..272...39W, Laigle2016ApJS..224...24L}), completeness, and consistency with our profile-fit WISE photometry. Details of the comparison and mass derivation methods are provided in Appendix \ref{sec:masses}. We therefore used \citetalias{Wen2024ApJS..272...39W} stellar masses to fit a unified mass–light–color relation calibrated separately for each photometry source (Table \ref{tab:coeff}):

$$\log M_*/M_\odot = \log D^2 +a \cdot z_\mathrm{mag} + b \cdot r_\mathrm{mag} + c \cdot W1_\mathrm{mag} + d \cdot z_\mathrm{ph} + e.$$

\begin{table}[]
\caption{Coefficients used for stellar mass estimation.}
    \centering
    \resizebox{0.8\columnwidth}{!}{

\begin{tabular}{lrrrrrr}
\hline
Photometry & a & b & c & d & e \\
\hline
LS DR9, DR10 & $-$0.81 & 0.56 & $-$0.24 & $-$0.5 & 13.1 \\
DELVE & $-$0.73 & 0.48 & $-$0.26 & $-$0.5 & 13.6 \\
PS1 & $-$0.64 & 0.43 & $-$0.29 & $-$0.5 & 13.7 \\
\hline
\end{tabular}
}
    
    \label{tab:coeff}
\end{table}

\begin{figure}
    \centering
    \includegraphics[width=0.8\linewidth]{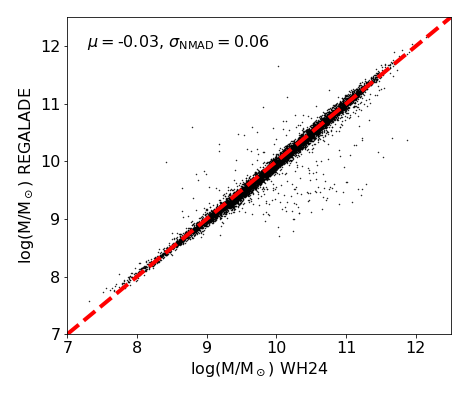}
    \caption{Comparison between our stellar mass estimates and the \citetalias{Wen2024ApJS..272...39W} reference values, for galaxies with distance estimates agreeing within 15\% in both catalogs. The legend reports the mean difference and normalized median absolute deviation (NMAD), in dex.}
    \label{fig:compa_masses}
\end{figure}

Allowing \(d\) to vary freely did not improve the fit; therefore, we fixed it to $-$0.5 based on the LS dataset. Removing this term, however, would significantly worsen the fit quality. Figure \ref{fig:compa_masses} shows a comparison of stellar masses in \citetalias{Wen2024ApJS..272...39W} to the obtained stellar mass estimates: the typical uncertainty is just $0.14$\,dex and the normalized median absolute deviation is $0.06$\,dex.

\subsection{Visual inspection of nearby galaxies}
\label{sec:classifgal}

\begin{figure}
 \centering
 \includegraphics[width=\linewidth]{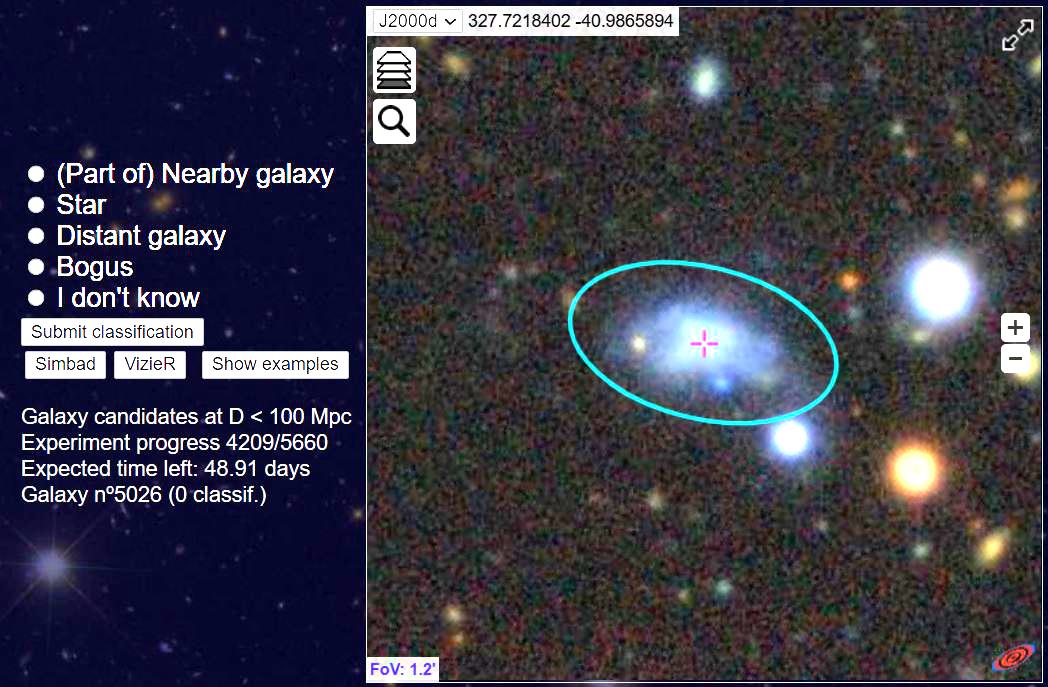}
 \caption{Screenshot of the galaxy classification platform, for a galaxy candidate at RA = 327.726$^\circ$, Dec = $-$40.995$^\circ$. The background is a ($g,r,i,z$) composite from Legacy Surveys DR10.}
 \label{fig:classifgal}
 \end{figure}

 Nearby galaxies (defined in this Section as $D<100$ Mpc) should appear in multiple input datasets due to their proximity and generally follow a predictable size trend, being large enough to match their close distances. However, some distant galaxies are catastrophic outliers from photometric redshift estimates, making them appear incorrectly within this range. To improve the purity of REGALADE in this range, we visually classified a random subset of 9,505 uncertain galaxy candidates at $D<100$\,Mpc having less than four distance estimates. For this purpose, we developed a dedicated online platform that collected approximately 27,000 independent classifications from 21 experts (Figure \ref{fig:classifgal}), with each object shown until it had received three independent evaluations. For each candidate, the image shown corresponded to the deepest available among the Aladin sky maps from Pan-STARRS, Legacy Surveys DR10, the DECam Plane Survey \citep{decaps2018ApJS..234...39S}, or SkyMapper \citep{Onken2024PASA...41...61O}. Each source was categorized into one of four classes: "(part of) nearby galaxy," "star," "distant galaxy," or "bogus," with an additional "I don't know" option. This process led to the derivation of the following refined constraints to clean the sample in this distance range ($D<100$\,Mpc), and to reduce the list of ambiguous cases requiring visual inspection:

 \begin{enumerate}
 \item Sources with a primary distance are retained without inspection due to their reliability.
 \item Sources with three or more distance estimates are considered reliable nearby galaxies.
 \item As introduced in Section \ref{sec:ps1}, the extent indicator $r-r_k$ helps distinguish nearby galaxies from stars or very distant galaxies. More than 90\% of objects with $r-r_K<0.9$ were classified as contaminant.
 \item sources without an optical counterpart are classified as spurious.
 \item Galaxies with $D<100$\,Mpc, $D_\mathrm{tmean}>200$\,Mpc are identified as distant.
 \item Groups of sources within 30 arc minutes containing more than 30 objects typically arise from inconsistent redshifts in deep, small-area surveys.
 \item Sources matching a Gaia object having $G<13$\,mag are identified as diffraction spike artifacts.
 \item Sources absent from all catalogs n$^\circ 0-9$ in Table \ref{tab:ranking} are treated as distant galaxies.
 \end{enumerate}

 To maximize the purity of the sample of galaxies at $D<100$\,Mpc, we visually classified the 4,383 sources having ($D<100$\,Mpc, $N_D<3$) but not following any of the above criteria.

 Sources receiving three independent classifications were finalized and removed from further inspection. Any object classified at least once as "(part of) nearby galaxy" was included in the final REGALADE catalog. A new parameter, "frac\_nearby," records the fraction of such classifications for each source. Sources classified as star or bogus by all annotators were removed from the sample. This discarded 1852 candidates.

\section{Results}\label{sec:results}

\subsection{Catalog at D<100 Mpc}

\begin{figure}
 \centering
 \includegraphics[width=\linewidth]{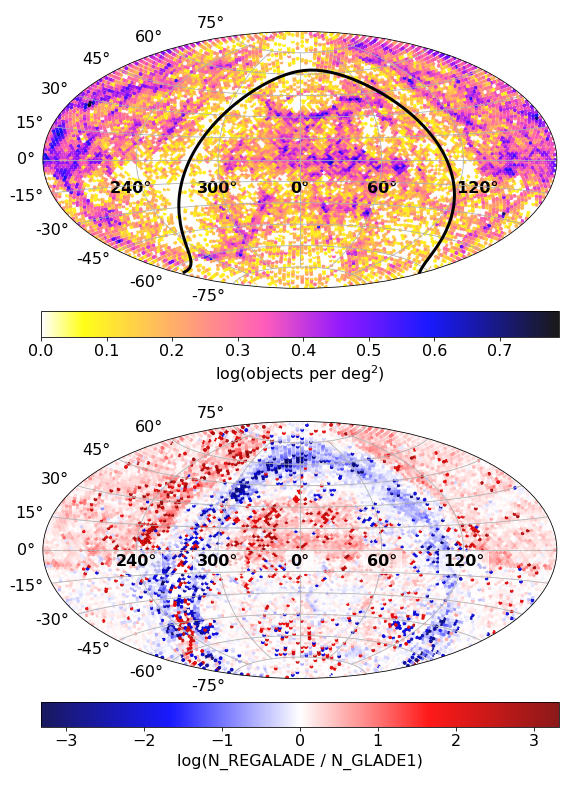}
 \caption{ (Top) Spatial distribution of REGALADE galaxies at $D<100$\,Mpc, in equatorial frame. The black line marks the Galactic plane. (Bottom) Differences in spatial density between REGALADE and GLADE1 galaxies at $D<100$\,Mpc (REGALADE excess in red, GLADE1 excess in blue).}
 \label{fig:sky_100}
 \end{figure}

The cleaned sample within 100\,Mpc includes 94,794 galaxies, representing a 55\% increase compared to HECATE and a 30\% increase relative to GLADE1. Figure \ref{fig:sky_100} illustrates the spatial distribution of REGALADE and GLADE catalogs within this distance range.

Superclusters are clearly visible in the compilation, including several structures whose coverage is largely provided by DESI DR1 and absent from other major galaxy catalogs. In contrast, GLADE1 shows a pronounced clustering near the Galactic plane ($|b|<15^\circ$), a region where many objects in that catalog are often flagged as contaminants in our compilation. This is also the case when comparing REGALADE to GLADE+. In both cases, REGALADE is more complete in large regions outside the Galactic plane, particularly in the northern hemisphere.

\subsection{Catalog overview}

Table \ref{tab:summary_metrics} describes the number of REGALADE galaxies in different subsets. Compared to any input catalog, the increase in sky coverage, number of galaxies with sizes, and spectroscopic redshifts is notable. Of great importance is the addition of stellar masses for 88\% of galaxies, particularly in the dwarf galaxy regime where GLADE+ misses most of their stellar mass estimates. 
Appendix \ref{sec:ap2} provides a description of REGALADE's content. 

\begin{table}[]
\caption{Summary of key metrics in the REGALADE catalog.}
\centering
\begin{tabular}{lr}
\hline
Metric & Value \\ \hline
Total number of galaxies & 79,880,104 \\
Sky coverage & 98.1 \% \\
Fraction with primary distance & 8.3\% \\
Fraction with $N_D \geq 2$ & 70.8\% \\
Median distance $D$ & 1260\,Mpc \\
Median $r_{K}$ magnitude & 20.2 \\
Median stellar mass $\log M/M_\odot$ & 9.9 \\
Number of galaxies with $R_1 > 5''$ & 19,878,606 \\
\hline
\end{tabular}

\label{tab:summary_metrics}
\end{table}

\begin{figure}
    \centering
    \includegraphics[width=\linewidth]{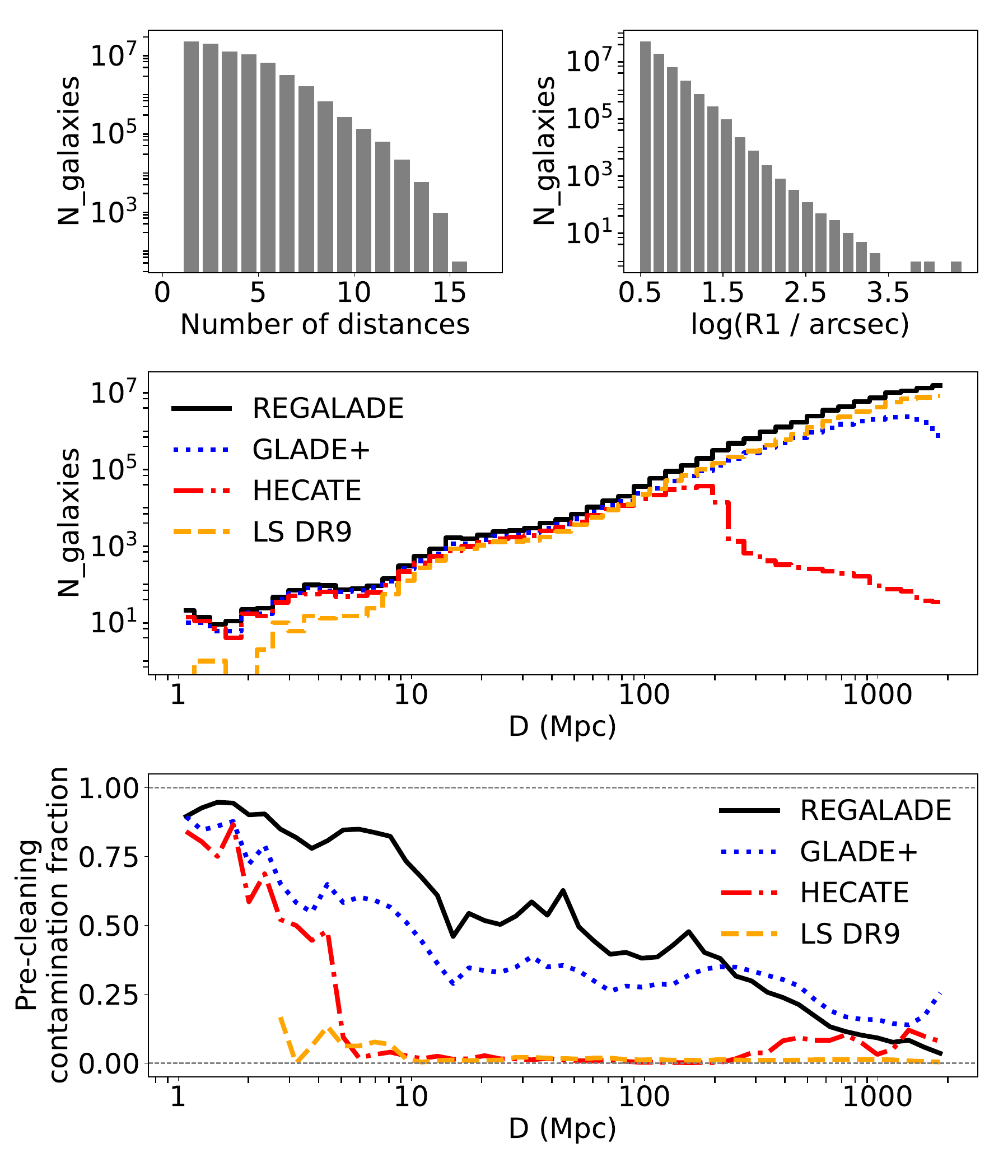}
    \caption{Distributions of key properties in the final REGALADE catalog.
Top left: Number of available distance estimates.
Top right: Semimajor axis $R_1$ (arcsec).
Middle: Distance distributions for REGALADE and selected input catalogs.
Bottom: Contamination fraction $N_{\rm contam}/(N_{\rm galaxies}+N_{\rm contam})$, calculated from the initial number of objects in each catalog before applying the cleaning process (i.e., the fraction subsequently removed).}
    \label{fig:disthist}
\end{figure}

Figure \ref{fig:disthist} illustrates the distribution of key quantities within REGALADE, including the number of matched distances per galaxy, the semimajor axis $R_1$, and the optimal distance estimate $D$. In particular, we highlight the contributions from several individual catalogs: while GLADE+ and HECATE contribute a larger number of galaxies at distances $D < 100$\,Mpc, LS galaxies dominate at greater distances. In all catalogs, contaminants are most prominent at low distances, with pre-cleaning contamination fractions above 30\% for REGALADE and GLADE+ at $D<100$\,Mpc.

\subsection{Distance accuracy}
\label{sec:dist_acc}

\begin{figure}
    \centering
    \includegraphics[width=\linewidth]{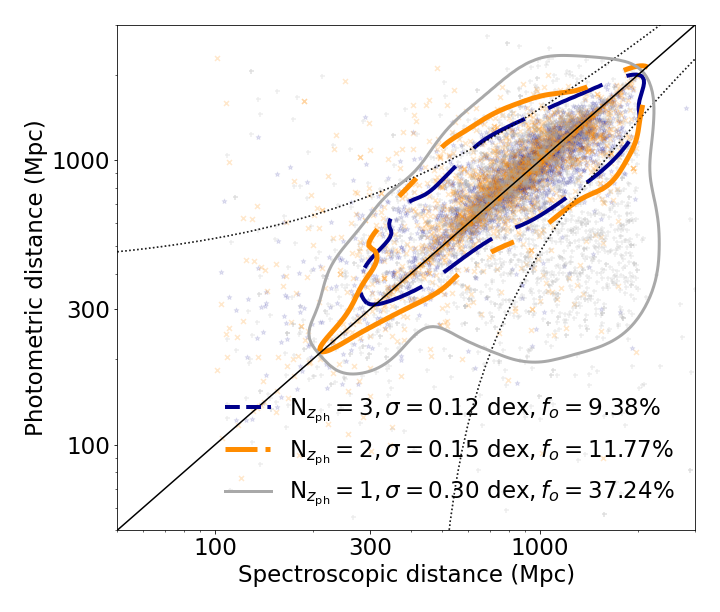}
    \caption{Mean photometric distance estimates against distances derived from DESI DR1 or NED spectroscopic redshifts, as a function of the number of available photometric redshift measurements ($\mathrm{N}_{z_\mathrm{ph}}$). Contours encompass 90\% of each distribution. Having multiple photometric redshifts available allows to derive more accurate distance estimates by effectively removing a significant fraction of outliers ($f_o$, defined as $|z_\text{sp} - z_\text{ph}| / (1 + z_\text{sp}) > 0.1$ and highlighted by the dotted lines) and by reducing the overall scatter in the distribution.}
    \label{fig:zph_ndist}
\end{figure}

Figure \ref{fig:zph_ndist} shows the mean photometric distance estimates plotted against spectroscopic distances derived from DESI DR1 or NED, categorized by the number of available photometric distance measurements ($\mathrm{N}_{z_\mathrm{ph}}$). This analysis focuses on galaxies having $g_K$, $r_K$, $i_K$, and $z_K$ brighter than 20\,mag. We define the outlier fraction, $f_o$, as the proportion of galaxies with a normalized redshift difference $|z_\text{sp} - z_\text{ph}| / (1 + z_\text{sp}) > 0.1$. These outliers largely drive the observed standard deviation ($\sigma$) when only one photometric estimate is available. The comparison shows that combining multiple independent datasets significantly improves the accuracy of REGALADE distance estimates. Since more than 70\% of REGALADE galaxies have $\mathrm{N}_{D}\geq 2$ (Table~\ref{tab:summary_metrics}), a large fraction of the catalog benefits from this effect.

\subsection{Completeness}

\begin{figure}
    \centering
    \includegraphics[width=0.9\linewidth]{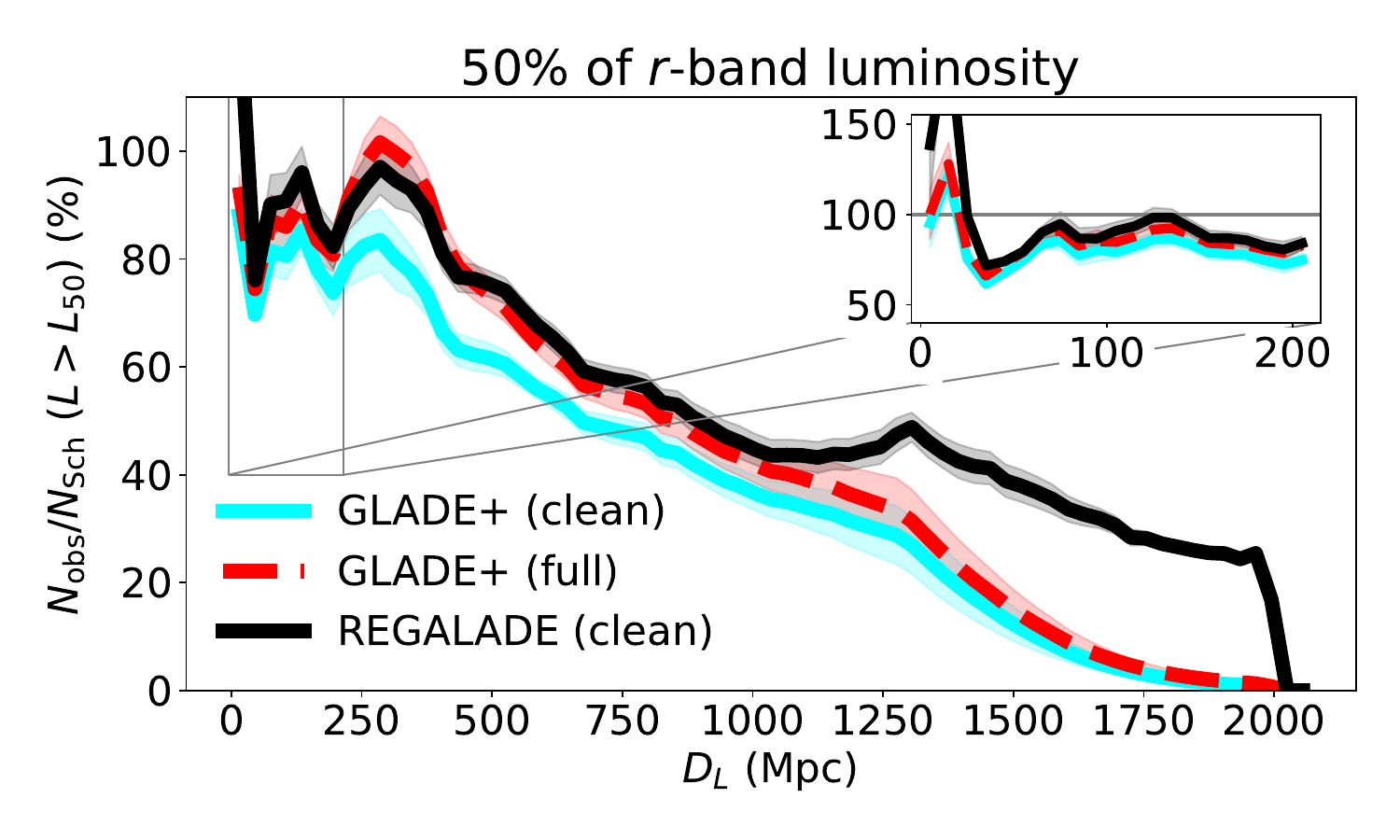}
    \includegraphics[width=0.9\linewidth]{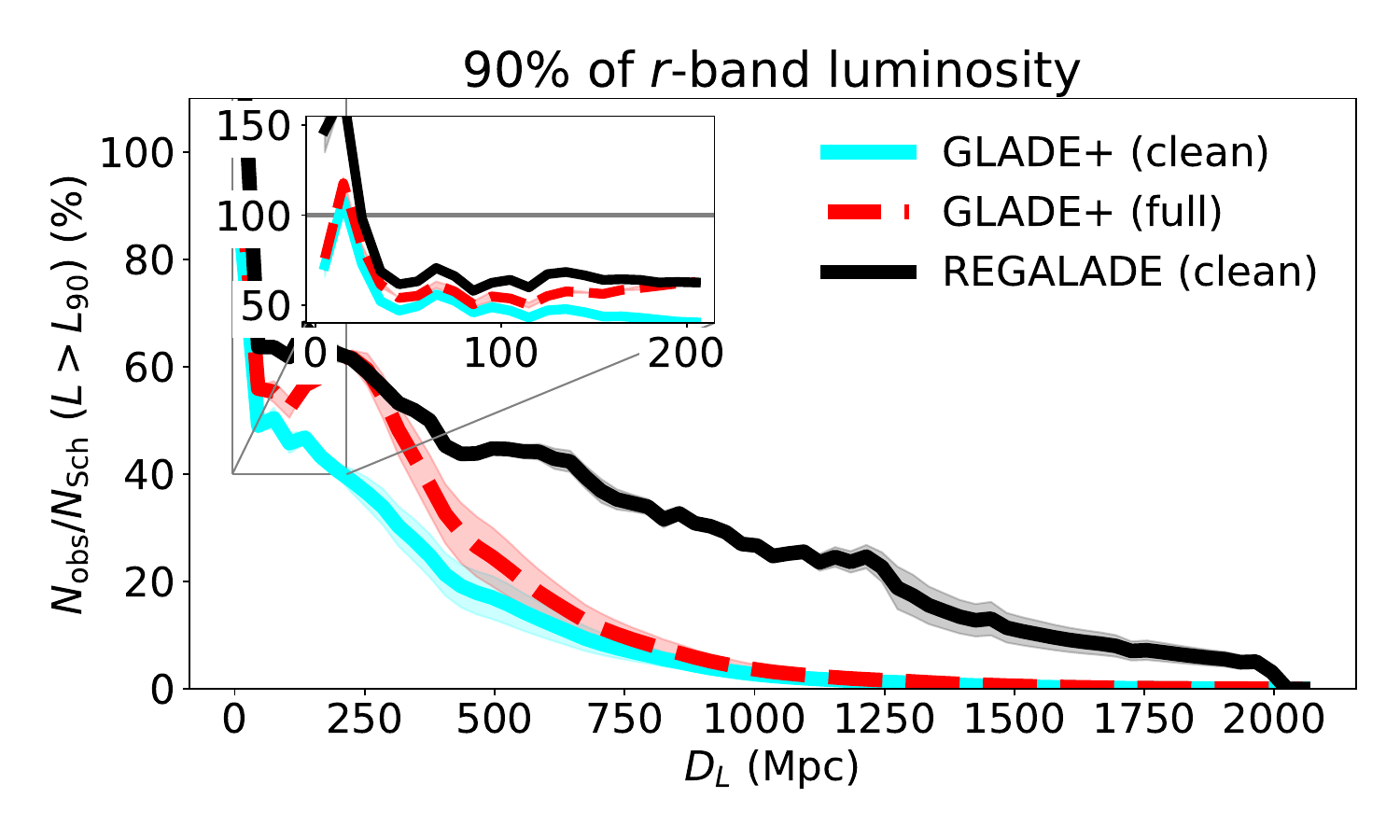}
    \caption{Completeness of REGALADE (black) and GLADE+, shown with contaminants (dashed red) and after cleaning (cyan), relative to the $r$-band Schechter luminosity functions of \citet{Blanton2003} and \citet{Hill2011}. Top panel: Completeness for galaxies contributing 50\% of the total $r$-band luminosity. Bottom panel: Completeness for galaxies contributing 90\% of the total $r$-band luminosity. The shaded region reflects the variation between the two luminosity function models.}
    \label{fig:completeness}
\end{figure}

Figure \ref{fig:completeness} shows the completeness of REGALADE and GLADE+ in 30 Mpc-wide distance shells (10 Mpc-wide in the range $0-200$\,Mpc, following the method of \citet{Dalya2018MNRAS.479.2374D} but using $r$-band luminosities. We adopted two Schechter function models derived from SDSS data -- \citet{Blanton2003}, based on Petrosian magnitudes, and \citet{Hill2011}, based on SDSS model magnitudes -- and integrated each above the luminosity threshold that contributes half of the total $r$-band light: $(L/L_*) = 0.645$ and $0.558$, respectively. This threshold corresponds to an $r$-band luminosity of $L_* \approx 2 \times 10^{10},L_\odot$. For typical stellar mass-to-light ratios of $M/L_r \sim 2$–$3$, this translates to a stellar mass of $\sim 5 \times 10^{10} M_\odot$, comparable to Milky Way-mass galaxies. The completeness estimate therefore applies to the population of such massive galaxies.

Legacy Survey magnitudes were converted to the SDSS system using the transformations in Appendix A of \citet{Dey2019AJ....157..168D}. K-corrections were applied using the $(g - r)$ color following \citet{Chilingarian2012}. To approximate SDSS Petrosian magnitudes, we applied a +0.06 mag offset to the Legacy model magnitudes before comparing them to the luminosity function of \citet{Hill2011}, consistent with the bias shown in their figure 13. Completeness was calculated as the ratio of galaxies above a luminosity threshold within each shell to the number expected based on the integrated Schechter function. Notably, completeness exceeds 100\% in the 10–20 Mpc range, indicative of a local overdensity of galaxies surrounding the Milky Way \citep{Dalya2018MNRAS.479.2374D}. In contrast, there is a significant dip in completeness near 40 Mpc, a feature that aligns with similar trends observed in the $B$-band and $K_S$-band luminosity functions for the GLADE and NED-LVS catalogs (see figure 4 of \citealt{Dalya2018MNRAS.479.2374D} and figure 11 of \citealt{Cook2023ApJS..268...14C}). Furthermore, a peak in the observed number of galaxies around 300 Mpc likely reflects the presence of numerous large-scale structures at this distance, such as the Sloan Great Wall \citep{Einasto2016A&A...595A..70E}, Pisces-Cetus supercluster \citep{Porter2005MNRAS.364.1387P}, and Horologium-Reticulum supercluster \citep{Fleenor2005AJ....130..957F}.

After cleaning, REGALADE demonstrates consistently higher completeness compared to GLADE+, even within the local Universe. Contaminants present in GLADE+ artificially inflate its completeness estimate by up to approximately 10\% out to distances of about 1000 Mpc. In contrast, REGALADE maintains greater than 90\% completeness out to roughly 360 Mpc.

The primary source of missed galaxies lies behind the Galactic plane, as completeness (in galaxies contributing half of the total $r$-band light) increases to nearly 100\% when considering only galaxies at Galactic latitudes $|b| > 20^\circ$. Conversely, in the complementary region ($|b| < 20^\circ$), completeness drops to $\sim$50\% within the 40--360 Mpc distance range. This decline underscores the challenges posed by Galactic extinction and source confusion near the plane.

\begin{figure}
\centering
\includegraphics[width=0.9\linewidth]{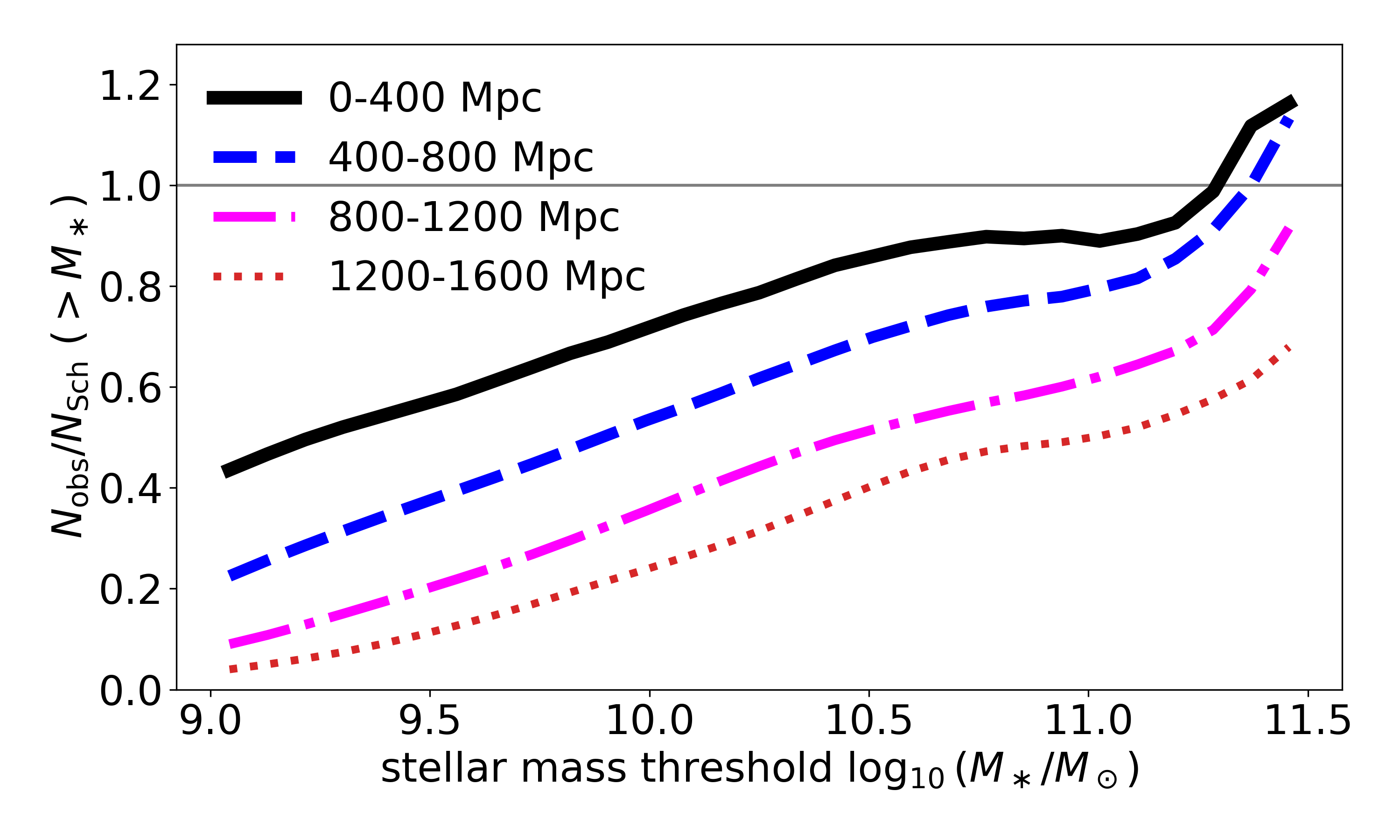}
\caption{Completeness of REGALADE with respect to the GSMF of \citet{Drory2009ApJ...707.1595D} as a function of the stellar mass threshold, shown for different distance bins. Photometric redshift uncertainties at the high-mass end can lead to overestimated number densities, resulting in apparent completeness values exceeding unity \citep{Drory2009ApJ...707.1595D}.}
\label{fig:gsmf}
\end{figure}

The bottom panel of Figure \ref{fig:completeness} highlights completeness for galaxies that contribute 90\% of the total luminosity, essentially representing the population of non-dwarf galaxies \citep{Staveley1992MNRAS.258..334S, Driver1999ApJ...526L..69D} with stellar masses $M_* \gtrsim 3 \times 10^9\,M_\odot$. Finally, Figure \ref{fig:gsmf} presents the completeness of REGALADE for various stellar mass thresholds, evaluated in separate distance bins. The reference is the galaxy stellar mass function (GSMF) of \citet{Drory2009ApJ...707.1595D} in their most local redshift bin $0.20 <z< 0.50$. At the high-mass end, photometric redshift uncertainties can artificially boost the inferred number densities, occasionally yielding completeness values above unity \citep[see][]{Drory2009ApJ...707.1595D}.

\section{Discussion}
\subsection{Applications of REGALADE}

We investigate the added value of REGALADE based on three different use cases: GW event target selection, optical transient host identification, and identification of luminous X-ray sources. The results are summarized in Table \ref{tab:sci_cases}.

\subsubsection{GW host candidates}

To demonstrate the potential of REGALADE in the search for electromagnetic counterparts to GW events, we crossmatched it with the current leading reference catalog, GLADE+, focusing on the confidence regions of two GW events: GW170817 and S250119cv. GW170817 is the only confirmed multi-messenger kilonova detected so far \citep{Abbott2017PhRvL.119p1101A}, while S250119cv represents a typical, well-localized binary black hole merger observed during the O4 run, located at approximately 500\,Mpc with a 90\% confidence region spanning 8.5 deg\(^2\). For the latter event, GLADE+ is claimed to be complete in terms of galaxies contributing 50\% of the  \(B\)-band total light \citep{Brozzetti2025GCN}.

Table \ref{tab:sci_cases} summarizes the number of galaxies from each catalog that lie within the 90\% confidence region of these events and their respective distance intervals (\(39.8 \pm 8.3\)\,Mpc for GW170817 and \(472 \pm 112\)\,Mpc for S250119cv). 
For the GW170817 event, host candidate counts are comparable, with minor differences attributable to distance estimates varying slightly between GLADE+ and REGALADE. Stellar masses for host galaxies tend to be on average 40\% higher in REGALADE, a result of incorporating \(W1\)-band photometry adapted to extended sources (see Section \ref{sec:photo}). Regarding S250119cv, the number of GLADE+ galaxies found here is lower than previously reported by \citet{Brozzetti2025GCN}, primarily because our crossmatch uses the intersection of the 90\% credible sky area and the 68\% confidence interval in distance for simplicity, unlike the 90\% credible volume considered in GLADEnet \citep{Brozzetti2024}. Nonetheless, REGALADE recovers approximately twice as many galaxies within the same search volume, largely due to improved completeness among dwarf galaxies. Indeed, host candidates included exclusively in REGALADE have a median stellar mass of about \(2 \times 10^9\,M_\odot\), roughly an order of magnitude lower than galaxies common to REGALADE and GLADE+.

\subsubsection{Transient host association}

To ensure transient-host completeness across different luminosity regimes, we analyzed two samples distinguished by luminosity: the 16,704 supernova-classified transients in the Transient Name Server (TNS, as of 25 May 2025) with published redshifts, and the transients reported to TNS by BlackGEM. BlackGEM is an array of optical telescopes situated at ESO's La Silla Observatory, capable of detecting transients down to approximately 22 magnitude. It is specifically designed for the rapid identification of kilonova signatures following GW events \citep{Groot}. Its Local Transient Survey targets extragalactic fields with significant mass concentrations, as well as galaxies within 15\,Mpc. This focus makes BlackGEM especially well-suited for discovering intrinsically low-luminosity transients, such as Iax supernovae and luminous red novae \citep{Groot}, while also providing complementary cadence for the upcoming LSST.

\begin{figure}
    \centering
    \includegraphics[width=0.9\linewidth]{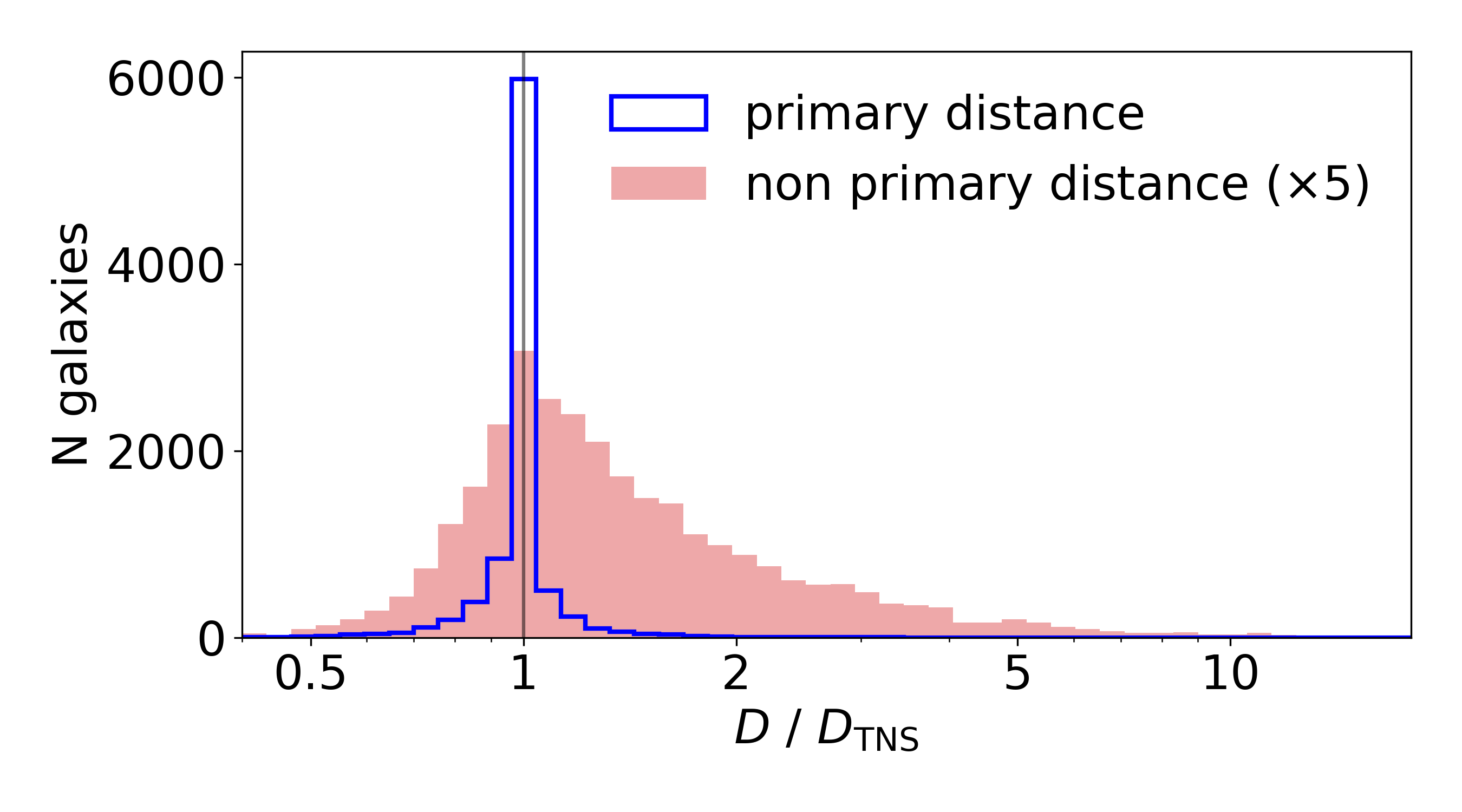}
    \caption{Ratio of REGALADE galaxy distances to those reported in TNS, for hosts with a primary distance estimate (blue) and a non-primary distance estimate (red; scaled by a factor of 5 for visibility). The narrow peak at unity for primary distances indicates strong agreement with TNS values and high reliability. Overall, distance accuracy is better than 30\% for 75\% of identified hosts, and within a factor of 2 for 90\% of them. The large number of transients with reliable host distances highlights the value of REGALADE for transient astronomy.}
    \label{fig:distcomp}
\end{figure}

The 16,704 TNS transients with published classifications are naturally biased toward high luminosity (half are discovered at $m_{\rm disc} < 19$,mag). Even within this bright subset, REGALADE successfully attaches a host to more than 90\% of events, recovering 15,101 galaxies within $1.5\times \mathrm{DLR}$\footnote{DLR being the directional light radius \citep{Sako2018PASP..130f4002S} defined from the galaxy's ellipse as the radius along the direction of the transient}, compared to 9,471 in GLADE1 -- a 60\% gain in completeness while retaining 99\% of the GLADE1 matches. The comparison between REGALADE and TNS distances for these hosts (Figure~\ref{fig:distcomp}) shows that most values are consistent, confirming the reliability of both the associations and the REGALADE distance estimates. 
The gain in completeness is even clearer for fainter transients: in the 1,605 TNS events discovered or codiscovered by BlackGEM (median $m_{\rm disc}=19.6$\,mag), REGALADE delivers 1,139 hosts, more than double the 507 in GLADE1. This is allowed by the inclusion of fainter and more distant galaxies, which account for a significant fraction of observed transients.

\subsubsection{X-ray source host association}

When we apply the quality cuts of \citet{Tranin_ulx}\footnote{Specifically, we select securely detected (maximum detection likelihood \texttt{SC\_DET\_ML}$>$10), point-like sources (extent value \texttt{SC\_EXTENT}=0 and extent likelihood \texttt{SC\_EXT\_ML}$<$100), with well-constrained position (\texttt{SC\_POSERR}$<$3'') and belonging to the clean sample (\texttt{SC\_SUM\_FLAG}$\leq$1). See \citet{Tranin_ulx}, Section 2.1.} to 4XMM-DR14 \citep{Webb2020}, a catalog of X-ray sources detected by the \textit{XMM-Newton} observatory, 52,390 point-like X-ray detections fall inside a REGALADE galaxy ellipse, compared with 24,753 for GLADE1. After removing nuclei and converting fluxes to luminosities with host distances, both catalogs yield a similar number of bona fide ULX hosts ($10^{39}-10^{41}$\,erg s$^{-1}$).\, The crucial difference is that one quarter of the GLADE1 associations are filtered out by REGALADE as Galactic objects or galaxies with overestimated distances. 
The contrast is sharper for HLX ($10^{41}-10^{43}$\,erg s$^{-1}$). REGALADE identifies 1159 HLX hosts -- 2.8 times more than GLADE1 after cleaning -- while discarding 27\% of the GLADE1 list as clumps of nearby galaxies or background quasars.

Across all tested science cases, REGALADE either recovers more legitimate hosts (up to a factor $\sim$2) or removes false positives that bias luminosity-dependent studies. The catalog therefore provides a more reliable base for GW target selection, transient follow-up, X-ray binary demographics, and any analysis that depends on accurate, volume-limited host statistics.

\begin{table}[ht]
\centering
\caption{Comparison of galaxy catalog performance across different science cases.}
\resizebox{\columnwidth}{!}{
\begin{tabular}{llrrr}
\hline
{Science Case} & {Catalog} & {N Galaxies} & {\% in common} \\
\hline
\multirow{2}{*}{GW170817 host candidates} & REGALADE & 28 & \\
& GLADE+ & 33  & 85\% \\
\hline
\multirow{2}{*}{S250119cv host candidates} & REGALADE & 1385  &  \\
& GLADE+ & 759  & 56\% \\
\hline
\multirow{2}{*}{TNS supernova hosts (16,704)} & REGALADE & 15,101 &  \\
& GLADE1 & 9,469  & 99.7\% \\
\hline
\multirow{2}{*}{BlackGEM transient hosts (1605)} & REGALADE & 1139  &  \\
& GLADE1 & 507 & 98.8\% \\
\hline
\multirow{2}{*}{ULX host (XMM)} & REGALADE & 1568  &  \\
& GLADE1 & 1422 & 73\% \\
\hline
\multirow{2}{*}{HLX host (XMM)} & REGALADE & 1159  &  \\
& GLADE1 & 566 & 73\% \\
\hline
\end{tabular}
}
\label{tab:sci_cases}
\end{table}

\subsection{Biases introduced by the compilation}

\begin{figure}
    \centering
    \includegraphics[width=\linewidth]{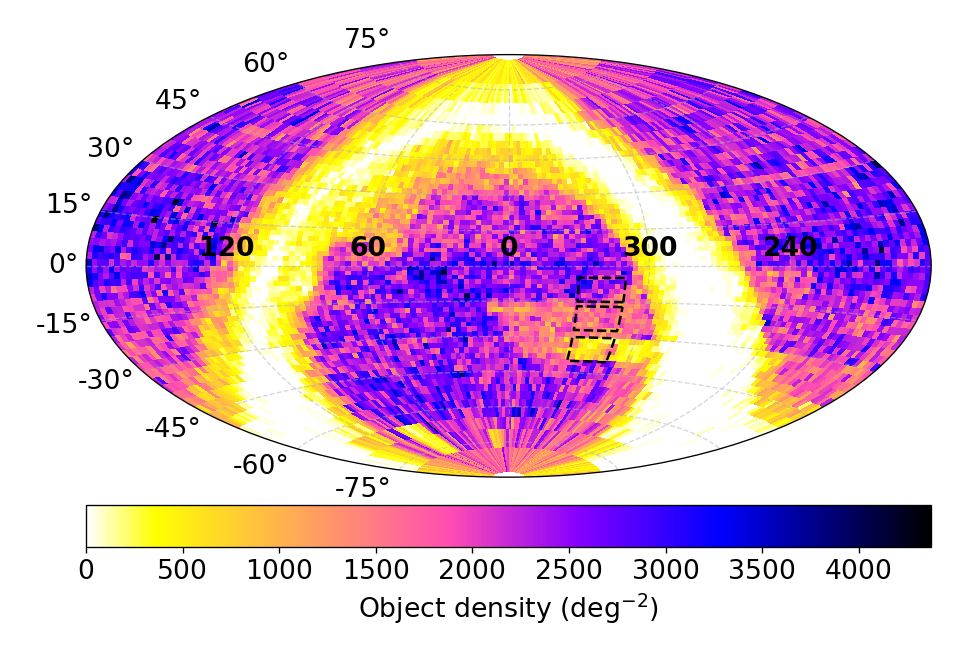}
    \caption{Aitoff projection of the sky distribution of REGALADE galaxies, with the color scale indicating the number density of galaxies per square degree. To illustrate variations in completeness, three fields characterized by different depths are highlighted.}
    \label{fig:aitoff_density}
\end{figure}

\begin{figure}
    \centering
    \includegraphics[width=0.9\linewidth]{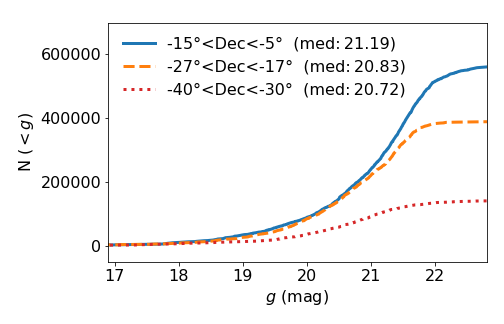}
    \caption{Cumulative distribution of $g_K$ magnitude in regions of different depth and similar area at $310<\mathrm{RA}<330$, shown in Figure \ref{fig:aitoff_density}.}
    \label{fig:hist_gk}
\end{figure}

Unifying multiple heterogeneous datasets comes with its challenges and surely leads to some issues in the resulting data. First of all, for being a heterogeneous selection, the selection function of REGALADE is unknown and completeness may vary greatly between different regions of the sky. Typically, objects behind the Galactic plane, in the zone of avoidance, are missed by most input catalogs, and if they only match one, they are removed at the contamination cleaning step (Section \ref{sec:contam}). Other regions of the extragalactic sky are covered at lower depth by the Legacy Surveys, such as the stripe between DECaLS and DES, at $300<\mathrm{RA}<330$ and $-40<\mathrm{Dec}<-15$. Figure \ref{fig:aitoff_density} shows the uneven sky distribution of REGALADE objects, with the locations of three representative fields of varying depth. To illustrate the impact of this variation on completeness, Figure \ref{fig:hist_gk} presents the cumulative $g_K$ magnitude distributions for three fields in the RA range [310$^\circ$, 330$^\circ$]: a deep DeCaLS field ($-15^\circ < \mathrm{Dec} < -5^\circ$), an intermediate-depth Pan-STARRS field ($-27^\circ < \mathrm{Dec} < -17^\circ$), and a shallow field ($-40^\circ < \mathrm{Dec} < -30^\circ$). Both the median magnitude and, more strikingly, the total number of detected galaxies decrease markedly in shallower regions. The forthcoming release of Legacy Survey DR11, with substantially expanded coverage toward the Galactic plane and improved depth in other areas, is expected to significantly mitigate these variations.

\subsection{Quasar retention}

Several of the cleaning steps applied to build REGALADE (Section \ref{sec:contam}) were designed to remove point-like sources and other likely contaminants.
However, such criteria inevitably risk excluding some genuine extragalactic sources with compact morphologies, particularly quasars. To quantify this effect, we crossmatched REGALADE with the high-purity, high-completeness catalog of mid-infrared–selected quasars of \citet{Secrest2015ApJS..221...12S}.

Restricting the comparison to low-redshift quasars ($z_{\mathrm{Secrest}} < 0.37$), we find that 272 objects are flagged as contaminants in REGALADE, while 11,300 are retained, corresponding to a loss fraction of only 2.3\%. Considering quasars of any redshift, 630 are removed versus 15,800 retained (3.8\% loss). For sources in the Secrest catalog without a measured redshift, 7,350 are excluded while 121,900 remain (5.6\% loss).

This relatively high retention rate is explained by the conservative filtering strategy adopted in Section~\ref{sec:contam}. Compact sources were removed only if they appeared in three or fewer of the input catalogs, a criterion that preserves well-known extragalactic objects. In addition, to avoid removing bona fide compact galaxies and quasars, we required that no match be found within a 5'' radius to a Legacy Survey source from \citet{Duncan2022MNRAS.512.3662D} with a low stellar probability ($p_{\rm star} < 0.5$). These criteria minimize the loss of quasars while still providing an effective reduction in stellar contamination. 

Overall, the lost fraction remains $\lesssim 5$\% in all cases, a reasonable trade-off given the significant improvement in catalog purity. Moreover, most of the excluded quasars lie beyond the design distance range of REGALADE.

 \subsection{Redshift completeness fraction}

The redshift completeness fraction (RCF), defined as the fraction of SN host galaxies with known spectroscopic redshifts prior to SN discovery, is an important metric for time-domain science \citep{Kulkarni2018ApJ...860...22K, Fremling2020}. It is expected to significantly improve in the coming years due to the ongoing and upcoming operations of large-scale spectroscopic surveys such as DESI and 4MOST. These surveys are targeting millions of galaxies across wide fields with deep limiting magnitudes, thus significantly expanding the pool of galaxies with secure spectroscopic redshifts. However, even more impactful in the near term is the increasing reliability and availability of high-quality photometric redshifts, especially for SN host galaxies. With multiple independent photo-z estimates and careful calibration using spectroscopic training sets, many galaxy catalogs now provide well-constrained photometric redshifts with small uncertainties, as demonstrated in Section \ref{sec:dist_acc}. These can be used for robust host identification. As a result, the effective completeness of SN host redshifts -- whether spectroscopic or high-confidence photometric -- is already much higher than traditional estimates suggest, and this enables a broader and more inclusive study of transient host environments using REGALADE-like catalogs, as demonstrated in the transient classification framework of \citet{Moller2020}. 

\begin{figure}
    \centering
    \includegraphics[width=0.8\linewidth]{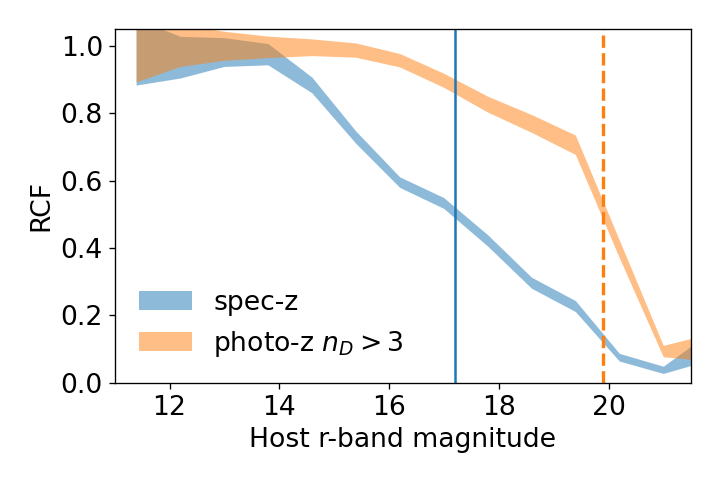}
    \caption{RCF of REGALADE SN hosts as a function of $r_K$ host magnitude, for spectroscopic and well-constrained photometric redshifts.}
    \label{fig:rband_rcf}
\end{figure}

In Figure \ref{fig:rband_rcf}, we replicate figure 5 from \citet{Fremling2020}, illustrating the RCF as a function of the Pan-STARRS $r_K$ magnitude for galaxies in our sample. REGALADE achieves a spectroscopic RCF of 0.5 at $r_K = 17.2$, marginally deeper than the limit reported by \citeauthor{Fremling2020} (16.9), even though our TNS-based sample of SN host galaxies tends to be less luminous and more distant compared to their ZTF Bright Transient Survey sample. When including hosts with at least four photometric redshift measurements, this completeness limit extends further to $r_K = 19.9$ mag.

 \subsection{Use of visual inspection data}
The visual classification of ambiguous sources within 100\,Mpc improves the reliability of the REGALADE sample and also provides a useful resource for the development of galaxy classification and photometric redshift algorithms. Unlike typical training sets based on spectroscopic samples or well-behaved sources, this dataset includes a realistic range of contaminants -- such as stars, artifacts, and parts of galaxies -- that are commonly found in wide-field photometric catalogs. This makes it well suited for testing and improving methods under conditions that more closely reflect those of real survey data. The inclusion of a classification confidence metric ("frac\_nearby") also allows for the development of probabilistic models. As photometric catalogs increasingly rely on automated classification and redshift estimation, training samples that incorporate this level of inspection and cleaning can help reduce systematic errors and improve overall performance.

Looking ahead, the visually vetted subset of REGALADE offers a foundation for developing automated classification and cleaning methods. Morphological classification projects such as Galaxy Zoo have shown the value of crowd-sourced, gold-standard labels for training machine-learning models on survey images (e.g., \citealt{HuertasCompany2011A&A...525A.157H,Dieleman2015MNRAS.450.1441D,Cheng2020MNRAS.493.4209C}). Similarly, the REGALADE visual classifications could be combined with multiband cutouts from Pan-STARRS and the Legacy Surveys to train deep-learning algorithms that better identify artifacts, stars, and fragmented galaxies. Future work could also explore a larger scale citizen-science component (e.g., through a Zooniverse project) to extend the training set and test the consistency of visual labels. Such efforts would help quantify label reliability and integrate human-verified classifications more directly into future releases of the catalog.

 \subsection{Upcoming data releases of deep surveys}

The next generation of deep imaging and spectroscopic surveys will greatly enhance the depth, completeness, and precision of galaxy compilations such as REGALADE. The forthcoming upGLADE catalog \citep{Brozzetti2024} will expand the existing GLADE+ compilation with improved coverage and deeper magnitude limits, greatly increasing completeness at distances $\gtrsim 1$\,Gpc. The upcoming Legacy Surveys DR11 will provide deeper optical and infrared imaging across regions currently lacking coverage, significantly extending the volume over which reliable host galaxy identification and distance estimation can be performed.

Meanwhile, future Euclid data releases will provide high-resolution, space-based NIR imaging and photometry across a large fraction of the extragalactic sky, complemented by photometric redshifts and morphological parameters. Euclid's combination of depth, wavelength coverage, and spatial resolution will be particularly powerful for identifying faint, low-surface-brightness galaxies and constraining their distances and stellar masses.

The Vera C. Rubin Observatory's LSST is expected to revolutionize transient science and galaxy catalogs alike. With its unprecedented depth, time-domain coverage, and photometric precision, LSST will not only discover billions of galaxies but also enable timely characterization of host galaxies for transients. 

These upcoming resources will enable a new era of galaxy compilations with much higher completeness at greater distances, improved purity through cross-survey validation, and better integration of photometric and morphological data. Combined with curated, low-redshift reference samples such as REGALADE, they will allow more accurate host galaxy association, population studies, and cosmological constraints.

\section{Summary and conclusion}\label{sec:disc_conc}

We presented REGALADE, a comprehensive compilation of galaxy catalogs focused on the nearby Universe ($D < 2000$\,Mpc), designed with an emphasis on purity, completeness, and enhanced value through inclusion of galaxy sizes, best and mean distances, profile-fit source magnitudes, and stellar mass estimates.

\begin{itemize}
    
\item To achieve completeness, we combined a variety of deep surveys covering the full sky (when combined) reaching typical point-source sensitivities fainter than 22 mag. REGALADE incorporates the most widely used nearby galaxy catalogs, using their well-defined sizes, redshift-independent distances, and both spectroscopic and photometric redshifts. This combination yields a final, clean sample of nearly 80 million galaxies that is $>90$\% complete for the brighter half of galaxy luminosities out to 360\,Mpc.
\item For purity, we used the Gaia catalog to effectively remove Milky Way stars, relied on the number of matched galaxy catalogs as a robust indicator of genuine galaxies, and applied selection cuts based on size, distance, and magnitudes to further eliminate false positives and residual stars. 
\item We emphasized the importance of using profile-fit W1 photometry for accurate stellar mass estimation, even for galaxies fainter than $W1 > 14$, which is sometimes considered negligible. Using Legacy Survey and corrected AllWISE W1 magnitudes, we derived precise stellar masses for 88\% of our catalog.
\item REGALADE is the first all-sky catalog at this depth to systematically include galaxy sizes and stellar masses. Furthermore, we demonstrated that combining multiple independent photometric redshift estimates significantly improves distance accuracy.
\item Our nearby sample ($D < 100$\,Mpc) was meticulously cleaned using sample cuts complemented by visual classification of all uncertain candidates.
\item We illustrated several applications of REGALADE, highlighting its capability to identify transient hosts, support electromagnetic counterpart searches for GW events, and enable discoveries of ULXs and HLXs.

\end{itemize}

The REGALADE preprocessed input and output catalogs are made available on Zenodo and/or through CDS. An interactive platform allows users to explore REGALADE galaxies, as well as the original input catalogs, via an Aladin Lite interface\footnote{\url{https://blackpearl.blackgem.org/regalade.php}} \citep{Bonnarel2000A&AS..143...33B, Boch2014ASPC..485..277B}. In addition, the REGALADE GitHub repository\footnote{\url{https://github.com/htranin/regalade}} will provide the codes required to reproduce this work and the demonstrated science applications.

\section{Data availability}
Table \ref{tab:subset} is only available in electronic form at the CDS via anonymous ftp to cdsarc.u-strasbg.fr (130.79.128.5) or via \url{http://cdsweb.u-strasbg.fr/cgi-bin/qcat?J/A+A/}.

\begin{acknowledgements}

We acknowledge financial support from grant CEX2024-001451-M, funded by MICIU/AEI/10.13039/501100011033. H.T. and N.B. acknowledge funding from the European Union (ERC, CET-3PO, 101042610). Views and opinions expressed are however those of the author(s) only and do not necessarily reflect those of the European Union or the European Research Council Executive Agency. Neither the European Union nor the granting authority can be held responsible for them. This work benefited from the contributions of numerous volunteers who participated in the visual classification of galaxy candidates. We thank the anonymous referee for its valuable comments, which significantly enhanced the quality of this paper.

This work made use of Astropy (\url{http://www.astropy.org}): a community-developed core Python package and an ecosystem of tools and resources for astronomy \citep{astropy:2013, astropy:2018, astropy:2022}. We acknowledge the usage of the TOPCAT astronomical software (\url{http://www.starlink.ac.uk/topcat/}).  This research made use of the crossmatch service provided by CDS, Strasbourg. This research has made use of the VizieR catalog access tool and the Aladin sky atlas, CDS, Strasbourg, France (DOI : 10.26093/cds/vizier). The original description of the VizieR service was published in  \citet{Vizier2000A&AS..143...23O}.

This research has made use of the NASA/IPAC Extragalactic Database (NED), which is operated by the Jet Propulsion Laboratory, California Institute of Technology, under contract with
the National Aeronautics and Space Administration. The description of NED-LVS can be found in \cite{Cook2023ApJS..268...14C} (DOI: 10.26132/NED8). We acknowledge the use of CasJobs and the JHU/ SDSS team for this powerful tool (\url{http://casjobs.sdss.org/CasJobs}). This research uses services or data provided by the Astro Data Lab, which is part of the Community Science and Data Center (CSDC) Program of NSF NOIRLab. NOIRLab is operated by the Association of Universities for Research in Astronomy (AURA), Inc. under a cooperative agreement with the U.S. National Science Foundation.

This research used data obtained with the Dark Energy Spectroscopic Instrument (DESI). DESI construction and operations is managed by the Lawrence Berkeley National Laboratory. This material is based upon work supported by the U.S. Department of Energy, Office of Science, Office of High-Energy Physics, under Contract No. DE–AC02–05CH11231, and by the National Energy Research Scientific Computing Center, a DOE Office of Science User Facility under the same contract. Additional support for DESI was provided by the U.S. National Science Foundation (NSF), Division of Astronomical Sciences under Contract No. AST-0950945 to the NSF’s National Optical-Infrared Astronomy Research Laboratory; the Science and Technology Facilities Council of the United Kingdom; the Gordon and Betty Moore Foundation; the Heising-Simons Foundation; the French Alternative Energies and Atomic Energy Commission (CEA); the National Council of Humanities, Science and Technology of Mexico (CONAHCYT); the Ministry of Science and Innovation of Spain (MICINN), and by the DESI Member Institutions: www.desi.lbl.gov/collaborating-institutions. The DESI collaboration is honored to be permitted to conduct scientific research on I’oligam Du’ag (Kitt Peak), a mountain with particular significance to the Tohono O’odham Nation. Any opinions, findings, and conclusions or recommendations expressed in this material are those of the author(s) and do not necessarily reflect the views of the U.S. National Science Foundation, the U.S. Department of Energy, or any of the listed funding agencies.

The DESI Legacy Imaging Surveys consist of three individual and complementary projects: the Dark Energy Camera Legacy Survey (DECaLS), the Beijing-Arizona Sky Survey (BASS), and the Mayall z-band Legacy Survey (MzLS). DECaLS, BASS and MzLS together include data obtained, respectively, at the Blanco telescope, Cerro Tololo Inter-American Observatory, NSF’s NOIRLab; the Bok telescope, Steward Observatory, University of Arizona; and the Mayall telescope, Kitt Peak National Observatory, NOIRLab. NOIRLab is operated by the Association of Universities for Research in Astronomy (AURA) under a cooperative agreement with the National Science Foundation. Pipeline processing and analyses of the data were supported by NOIRLab and the Lawrence Berkeley National Laboratory (LBNL). Legacy Surveys also uses data products from the Near-Earth Object Wide-field Infrared Survey Explorer (NEOWISE), a project of the Jet Propulsion Laboratory/California Institute of Technology, funded by the National Aeronautics and Space Administration. Legacy Surveys was supported by: the Director, Office of Science, Office of High Energy Physics of the U.S. Department of Energy; the National Energy Research Scientific Computing Center, a DOE Office of Science User Facility; the U.S. National Science Foundation, Division of Astronomical Sciences; the National Astronomical Observatories of China, the Chinese Academy of Sciences and the Chinese National Natural Science Foundation. LBNL is managed by the Regents of the University of California under contract to the U.S. Department of Energy. The complete acknowledgments can be found at \url{https://www.legacysurvey.org/acknowledgment/}. The Legacy Surveys Sky Viewer web sites \url{viewer.legacysurvey.org} and \url{legacysurvey.org/viewer} provide images from a number of surveys who have publicly released their data.

The Siena Galaxy Atlas was made possible by funding support from the U.S. Department of Energy, Office of Science, Office of High Energy Physics under Award Number DE-SC0020086 and from the National Science Foundation under grant AST-1616414. The DELVE Survey gratefully acknowledges support from Fermilab LDRD (L2019.011), the NASA Fermi Guest Investigator Program Cycle 9 (No. 91201), the National Science Foundation (AST-2108168, AST-2108169, AST-2307126, AST-2407526), and the NSF-Simonyi Scholars program.

Funding for the SDSS IV has been provided by the Alfred P. Sloan Foundation, the U.S. Department of Energy Office of Science, and the Participating Institutions. SDSS acknowledges support and resources from the Center for High-Performance Computing at the University of Utah. The SDSS web site is \url{www.sdss4.org}. SDSS is managed by the Astrophysical Research Consortium for the Participating Institutions of the SDSS Collaboration including the Brazilian Participation Group, the Carnegie Institution for Science, Carnegie Mellon University, Center for Astrophysics | Harvard \& Smithsonian (CfA), the Chilean Participation Group, the French Participation Group, Instituto de Astrofísica de Canarias, The Johns Hopkins University, Kavli Institute for the Physics and Mathematics of the Universe (IPMU) / University of Tokyo, the Korean Participation Group, Lawrence Berkeley National Laboratory, Leibniz Institut f\"ur Astrophysik Potsdam (AIP), Max-Planck- Institut f\"ur Astronomie (MPIA Heidelberg), Max-Planck-Institut f\"ur Astrophysik (MPA Garching), Max-Planck-Institut f\"ur Extraterrestrische Physik (MPE), National Astronomical Observatories of China, New Mexico State University, New York University, University of Notre Dame, Observatorio Nacional / MCTI, The Ohio State University, Pennsylvania State University, Shanghai Astronomical Observatory, United Kingdom Participation Group, Universidad Nacional Autonoma de Mexico, University of Arizona, University of Colorado Boulder, University of Oxford, University of Portsmouth, University of Utah, University of Virginia, University of Washington, University of Wisconsin, Vanderbilt University, and Yale University.

This work has made use of data from the European Space Agency (ESA) mission {\it Gaia} (\url{https://www.cosmos.esa.int/gaia}), processed by the {\it Gaia} Data Processing and Analysis Consortium (DPAC, \url{https://www.cosmos.esa.int/web/gaia/dpac/consortium}). Funding for the DPAC has been provided by national institutions, in particular the institutions participating in the {\it Gaia} Multilateral Agreement.

This publication makes use of data products from the Two Micron All Sky Survey, which is a joint project of the University of Massachusetts and the Infrared Processing and Analysis Center/California Institute of Technology, funded by the National Aeronautics and Space Administration and the National Science Foundation. This publication makes use of data products from the Wide-field Infrared Survey Explorer, which is a joint project of the University of California, Los Angeles, and the Jet Propulsion Laboratory/California Institute of Technology, and NEOWISE, which is a project of the Jet Propulsion Laboratory/California Institute of Technology. WISE and NEOWISE are funded by the National Aeronautics and Space Administration.

This research has made use of data obtained from the 4XMM \textit{XMM-Newton} Serendipitous Source Catalog compiled by the 10 institutes of the \textit{XMM-Newton} Survey Science Centre selected by ESA.

\end{acknowledgements}

\bibliographystyle{aa}
\bibliography{aanda}
\appendix
\normalsize

\section{SQL queries}

In the following we provide the SQL queries used to retrieve large galaxy catalogs.

\begin{enumerate}
    \item SDSS DR17 galaxies\footnote{\url{https://www.sdss4.org/dr14/algorithms/classify/}}
    
\texttt{SELECT ra, dec, Photoz.z as zph, zErr, devMag\_r, deVRad\_r, deVAB\_r, deVPhi\_r 
FROM Photoz JOIN Galaxy ON Photoz.objid = Galaxy.objid
WHERE clean = 1 AND deVMag\_g<21 AND deVRad\_r>2 AND Photoz.z>0
  AND  deVMag\_r<25-5*log10(deVRad\_r)}\\

    \item Pan-STARRS DR1 galaxies \citep{Chambers2016}
    
\texttt{SELECT 
        RAJ2000, DEJ2000, gmag, gKmag, rmag, rKmag, imag, iKmag, zmag, zKmag
        FROM "II/349/ps1"
        WHERE Ns>0 AND iKmag BETWEEN 13 AND 21 AND zmag>zKmag+0.3 AND gmag>gKmag+0.3}\\

    \item Pan-STARRS DR2 galaxy profiles \citep{Flewelling2020}
    
\texttt{SELECT o.raMean, o.decMean, m.rGalIndex, m.rGalMajor, m.rGalMinor, m.rGalPhi
    FROM ObjectThin o
    INNER JOIN ForcedGalaxyShape m ON o.objID = m.objID
    WHERE
        m.zGalMag between 13 and 20.5 and m.iGalMag<20.5
        AND (m.rGalIndex<2 AND m.rGalMajor>1.5 OR m.rGalIndex = 4 AND m.rGalMajor>2)}\\

    \item DELVE DR2 galaxies \citep{Drlica2022ApJS..261...38D}

\texttt{SELECT ra, dec, a\_image\_r, b\_image\_r, theta\_image\_r, mag\_auto\_g, mag\_auto\_r, mag\_auto\_i, mag\_auto\_z
FROM delve\_dr2.objects
WHERE mag\_auto\_r<21 AND extended\_class\_r>1}\\

    \item DELVE DR2 photometric redshifts \citep{Drlica2022ApJS..261...38D}

\texttt{SELECT ra, dec, z, zerr FROM delve\_dr2.photoz
WHERE z<0.5 AND extended\_class>1 
AND flag\_nband>3 AND flag\_sn10>2}\\

    \item GSC northern blue galaxies \citep{Lasker2008AJ....136..735L}

\texttt{SELECT 
        RA\_ICRS, DE\_ICRS, a, e, PA
        FROM "I/353/gsc242"
        WHERE "gmag"<22.3 AND rmag<22. AND imag<21.8 
        AND rmag-imag<0.5 AND ("gmag"-rmag<0.8 OR W1mag IS NOT NULL) 
        AND "Gmag" IS NULL
}\\

    \item GSC bright blue galaxies \citep{Lasker2008AJ....136..735L}

\texttt{SELECT 
        RA\_ICRS, DE\_ICRS, a, e, PA
        FROM "I/353/gsc242"
        WHERE Bjmag IS NOT NULL AND (Bjmag<21.5 OR W1mag IS NOT NULL) AND  "Gmag" IS NULL
        }
    
\end{enumerate}

\label{sec:sql}

\section{Input catalog homogenization}
\label{sec:ap1}
\begin{figure}
    \centering
    \includegraphics[width=0.8\linewidth]{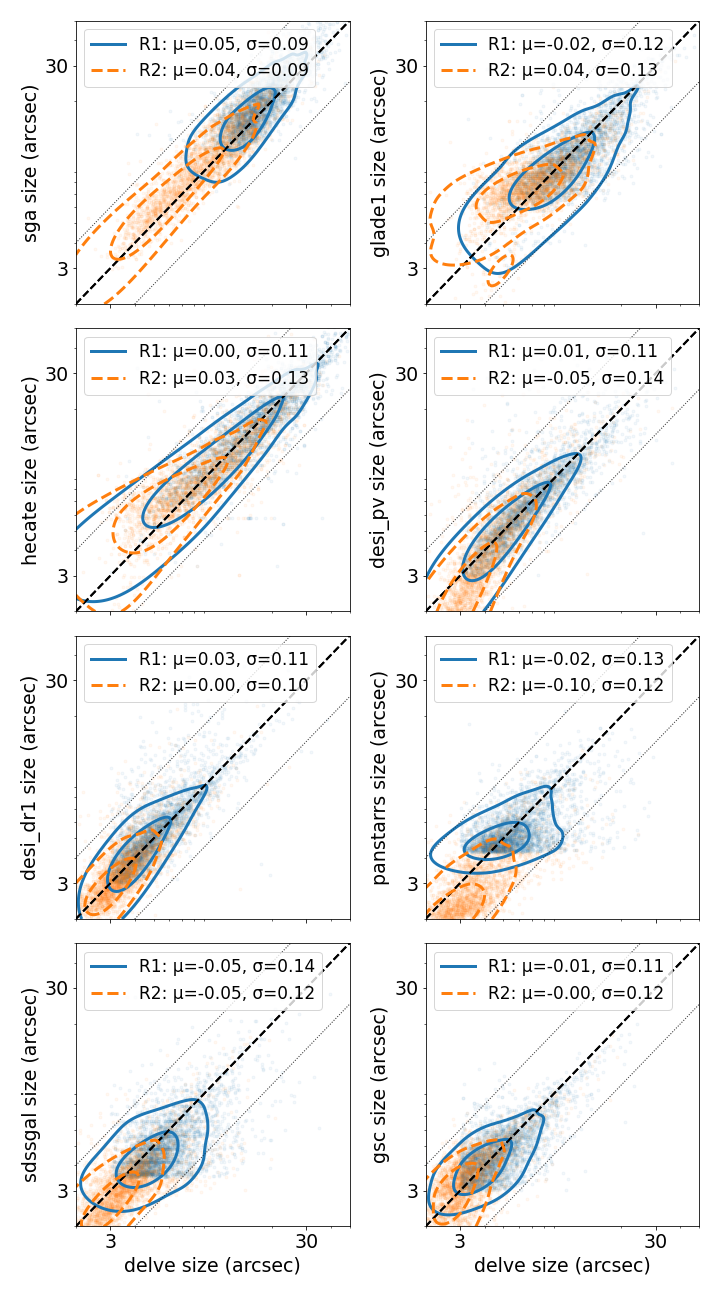}
    \caption{Comparison between sizes measured from different catalogs, after calibration. Both semimajor and semiminor axes are shown and compared to the DELVE reference. The legend gives the mean bias and standard deviation between the measurements, in dex. Dotted lines represent a deviation of a factor 2. Contours encompass 50\% and 90\% of the distribution.}
    \label{fig:r1}
\end{figure}

Figure \ref{fig:r1} compares the parameters ($R_1$,$R_2$) of galaxies across various input catalogs (here matched using a 3 arcsec radius for secure matches). SGA provides precise measurements of galaxy sizes, thanks to the depth of the Legacy Surveys and a rigorous ellipse-fitting methodology. The larger DELVE catalog -- also based on Legacy Surveys -- and its Sextractor-derived sizes show excellent agreement with SGA, so we can use it as a reference for plotting. Galaxy sizes measured from LS images tend to be larger than those found in classical catalogs such as GLADE1, HECATE and HyperLEDA, precisely because they resolve the outskirts of galaxies down to lower surface 
brightness, so the latter sizes are increased by 30\% (Table \ref{tab:calib}). This comparison motivated the choice of the "best ellipse" among available ellipses in the REGALADE compilation (Table \ref{tab:ranking}).

\begin{table}[]
\caption{Equations used for size calibration}
    \centering
    \resizebox{\columnwidth}{!}{
    \begin{tabular}{ccc}
    \hline
    Catalog & Input parameters $p_1, p_2, p_3$& Equations\\\hline
            &        \texttt{a\_image\_r},             &  $R_1 = p_1$ \\
    DELVE   &   \texttt{b\_image\_r},                  &  $R_2 = p_2$\\
            &        \texttt{theta\_image\_r}          &  $PA = 180 - p_3$ \\\hline
            &                                          &  $R_1 = 30~ p_1$ \\
    SGA     &  \texttt{D26}, \texttt{BA}, \texttt{PA}  &  $R_2 = 30~ p_1p_2$\\
            &                                          &  $PA = p_3$ \\\hline
            &                                          &  $R_1 = 1.3\cdot 3\cdot 10^{p_1}$ \\
  HyperLEDA &  \texttt{logD25}, \texttt{logR25}, \texttt{PA}  &  $R_2 = 1.3\cdot 3\cdot 10^{p_1-p_2}$\\
            &                                          &  $PA = p_3$ \\\hline
            &                                          &  $R_1 = 1.3\cdot 60~ p_1$ \\
    HECATE  &  \texttt{R1}, \texttt{R2}, \texttt{PA}    &  $R_2 = 1.3\cdot 60~ p_2$\\
            &                                          &  $PA = p_3$ \\\hline
            & (\texttt{R1\_GWGC}, \texttt{R1\_Hyp}, \texttt{R1\_2MASS}), &  $R_1 = 1.3\cdot \max(p_1)$ \\
    GLADE1  & (\texttt{R2\_GWGC}, \texttt{R2\_Hyp}, \texttt{R2\_2MASS}), &  $R_2 = 1.3\cdot \max(p_2)$\\
            & (\texttt{PA\_GWGC}, \texttt{PA\_Hyp}, \texttt{PA\_2MASS}) &  $PA = \mathrm{coalesce}(p_3) $ \\\hline
            &                                          &  $R_1 = 2.5~ p_1+1.5$ \\
   DESI DR1 &  \texttt{SHAPE\_R}, \texttt{SHAPE\_E1}, \texttt{SHAPE\_E2}    &  $R_2 =R_1\big(1-(p_2^2+p_3^2)^{1/4}\big)$\\
            &                                          &  $PA = \begin{cases}\tan^{-1}(p_3/p_2) &p_2>0 \\ \tan^{-1}(p_3/p_2)+90 &p_2<0\end{cases} $ \\\hline
            &                                          &  $R_1 = 2.5~ p_1+1.5$ \\
    DESI PV &  \texttt{ShapeR}, \texttt{b/a}, \texttt{PA} &  $R_2 = p_2(2.5~  p_1+1.5)$\\
            &                                          &  $PA = p_3$ \\\hline
            &                                          &  $R_1 = 2.5~ p_1^\alpha+0.5$ \\
            &                                          &  $R_2 = 2.5~ p_2^\alpha +0.5$\\
  Pan-STARRS &  \texttt{rGalMajor}, \texttt{rGalMinor}, \texttt{rGalPhi}                    &  $PA = p_3+90$ \\
            &                                          &  $\alpha =\begin{cases}1& \text{Sersic model}\\0.6& \text{de Vaucouleurs model} \end{cases}$\\\hline
            &                                          &  $R_1 = p_1+1.5$ \\
    SDSS    &  \texttt{deVRad\_r}, \texttt{deVAB\_r}, \texttt{deVPhi\_r}    &  $R_2 = p_2+1.5$\\
            &                                          &  $PA = p_3$ \\\hline
    \end{tabular}
    }
    
    \label{tab:calib}
\end{table}

\begin{figure}
    \centering
    \includegraphics[width=0.8\linewidth]{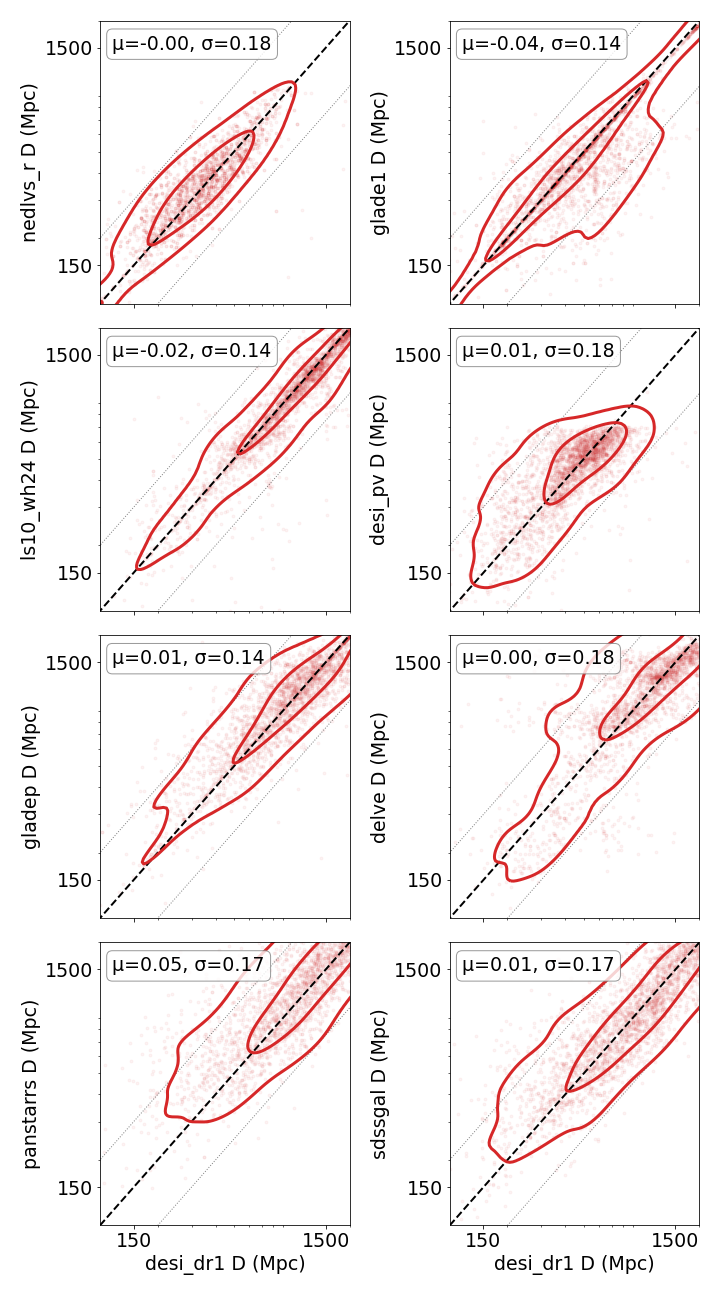}
    \caption{Comparison between distances measured from different catalogs. Redshift-based distances are shown and compared to the DESI DR1 reference. The legend gives the mean bias and standard deviation between the measurements, in dex. Dotted lines represent a deviation of a factor 2. Contours encompass 50\% and 90\% of the distribution.}
    \label{fig:d}
\end{figure}

\begin{figure}
    \centering
    \includegraphics[width=0.9\linewidth]{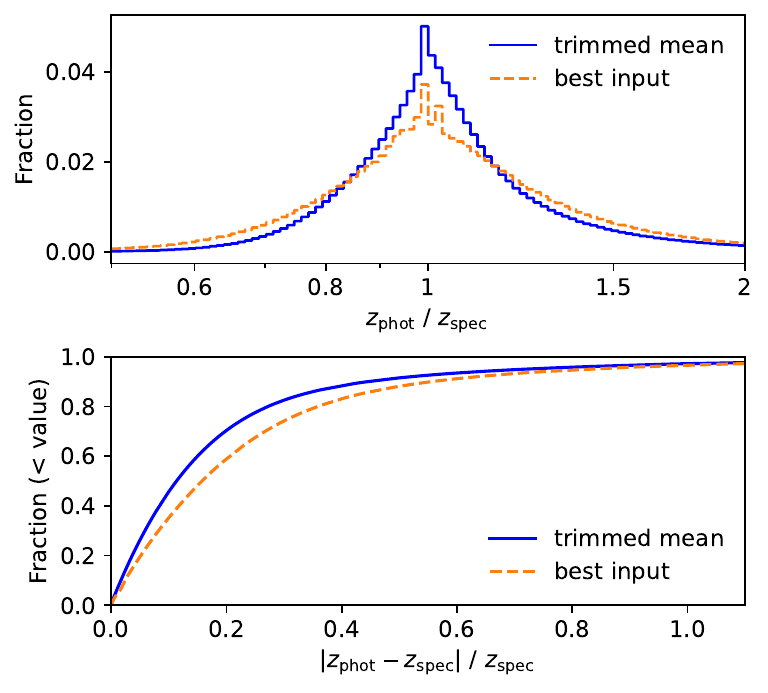}
    \caption{Comparison between photometric and spectroscopic redshifts for galaxies with available $z_{\mathrm{spec}}$.
(Top) Distribution of the ratio between the photometric redshift from the preferred catalog ("best input," orange) or the trimmed mean of photometric estimates (blue) and the spectroscopic redshift.
(Bottom) Cumulative distribution of the relative difference $|z_{\mathrm{phot}} - z_{\mathrm{spec}}|/z_{\mathrm{spec}}$.}
    \label{fig:dratio}
\end{figure}

Similarly, we compare the photometric distance estimates from the input catalogs in Figure \ref{fig:d} with the DESI DR1 distances based on spectroscopic redshifts for reference. As introduced in Section \ref{sec:dist_acc}, averaging multiple estimates can improve the accuracy of the distance estimate. In Figure \ref{fig:dratio}, we compare both the best input photometric redshift and the trimmed mean of photometric redshifts to the spectroscopic reference. The distribution using trimmed-mean redshifts is symmetric and shows a higher fraction of galaxies with small relative differences compared to the best input.

\section{WISE photometry correction}
\label{sec:w1ext_proxies}

 We identified a systematic discrepancy between WISE point-source photometry ($W1\_\mathrm{AllWISE}$) and the profile-fit values from Legacy Surveys ($W1\_\mathrm{LS}$), which becomes significant for extended galaxies. This bias arises because AllWISE measures point-source fluxes, while the Tractor software \citep{Dey2019AJ....157..168D} used in the Legacy Surveys models extended sources across optical and infrared bands.

To correct this bias, we calibrated empirical corrections using galaxies from HECATE overlapping the SDSS footprint, where profile-fit forced photometry $WF1$ is available. We trained a random-forest regressor to predict the correction $W1_\mathrm{corr} = WF1 - W1\_\mathrm{AllWISE}$ using features from REGALADE. The three most important features were $(W1\_\mathrm{LS} - W1\_\mathrm{AllWISE})$, $(K\_\mathrm{ext} - K)$, and $\sqrt{R_1}$. Based on these, we fitted linear empirical formulae to derive corrections:

$$
W1_\mathrm{corr} \approx \frac{K_\mathrm{ext} - K}{2}
\quad\text{(for 2MASS XSC sources)},
$$

$$
W1_\mathrm{corr} \approx 1 - 3.3\,\sqrt{R_1\,(\mathrm{arcmin})}
\quad\text{(for other sources with }R_1 > 5\arcsec).
$$

\begin{figure}
    \centering
    \includegraphics[width=\linewidth]{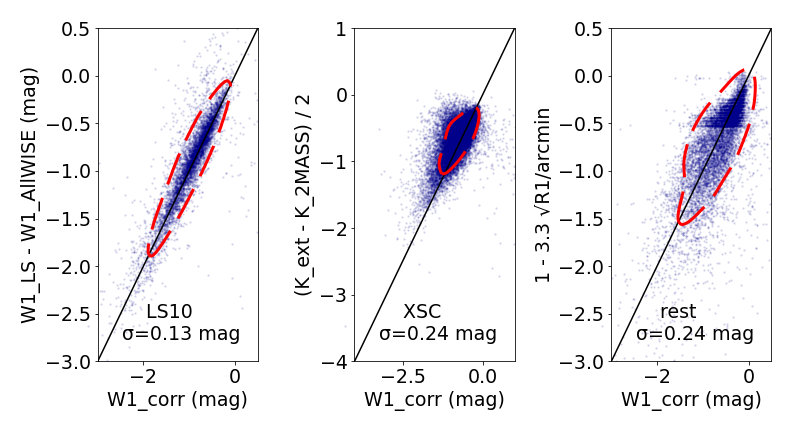}
    \caption{Relations used to correct WISE photometry of extended sources. (Left) LS \textit{Tractor} $W1$ magnitude minus AllWISE magnitude. (Middle) Rescaled 2MASS $K$-band XSC magnitude minus point-like magnitude. (Right) Correction based on galaxy radius, for those sources not in 2MASS XSC. The NMAD is given in the legend. Contours encompass 70\% of each distribution.}
    \label{fig:w1ext_proxies}
\end{figure}

Figure~\ref{fig:w1ext_proxies} compares the proxy corrections to the true corrections. Although previous work suggested that such corrections are negligible for $W1 > 13.8$ \citep{Bilicki2016ApJS..225....5B}, we find that 73\%, 77\%, and 5\% of galaxies in the three respective subsamples (with LS photometry, 2MASS XSC, or neither) are brighter than this threshold -- confirming the need to apply these corrections in most cases. 

\section{Stellar mass estimates}
\label{sec:masses}
Some of our input galaxy catalogs provide stellar mass estimates, including HECATE, GLADE+, LS DR9 galaxies from \citet{Zou2022RAA....22f5001Z}, and LS DR10 galaxies from \citetalias{Wen2024ApJS..272...39W}. A common approach estimates stellar masses via calibrated mass-to-light ratios, where the “light” is typically near-infrared (e.g., $W1$ or $K$ band) luminosity, and the reference masses are derived from stellar population synthesis fits to spectra or multiband photometry \citep[e.g.,][]{Bell2003ApJS..149..289B, Driver2009A&G....50e..12D}. These calibrations often relate mass-to-light ratios to optical or IR colors \citep{Cluver2014ApJ...782...90C, Kettlety2018MNRAS.473..776K}.

\cite{Parkash2018ApJ...864...40P} used profile-fit WISE photometry to estimate masses for HIPASS galaxies based on $W1$ luminosity and $W1-W2$ color. Similarly, \cite{Kovlakas2021MNRAS.506.1896K} computed HECATE stellar masses using $K$-band luminosities and $(g - r)$-based calibrations, benchmarking them against SED fits from the GALEX-SDSS-WISE catalog (GSWLC; \citealt{Salim2016ApJS..227....2S}). \cite{Zou2022RAA....22f5001Z} used machine-learned photometric redshifts to build $grzW1W2$ SEDs fitted with \cite{Bruzual2003MNRAS.344.1000B} models via Le Phare \citep{Laigle2016ApJS..224...24L}. NSA stellar masses from \cite{Blanton2007AJ....133..734B} follow a similar template-fitting method. These catalogs provide useful benchmarks for comparison.

\begin{figure}
\centering
\includegraphics[width=0.9\linewidth]{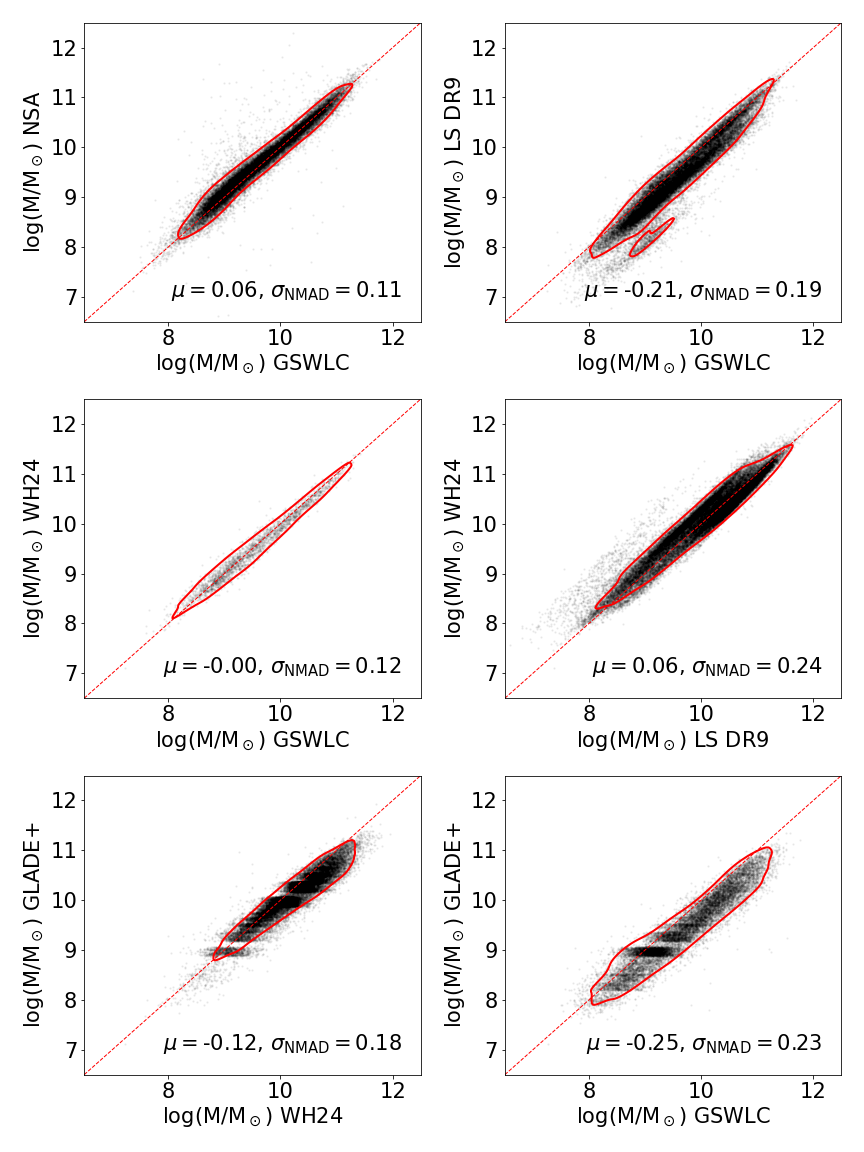}
\caption{Comparison between stellar masses of our input catalogs. 
Row 1: Template-fitting model estimates of GSWLC \citep{Salim2016ApJS..227....2S}, NASA Sloan Atlas \citep{Blanton2007AJ....133..734B}, and LS DR9 galaxies \citep{Zou2022RAA....22f5001Z}. Row 2: Photometry-based stellar masses (LS10 galaxies) against template-based estimates. Row 3: GLADE+ stellar masses against reference estimates. Contours encompass 90\% of each distribution. The mean difference and NMAD are quoted in legend.}
\label{fig:compa_mass_other}
\end{figure}

Figure \ref{fig:compa_mass_other} shows the consistency between these catalogs (after adjusting for different $H_0$ and matching distances within 5\%), with a typical scatter of 0.15 dex (NMAD). GLADE+ masses are systematically lower and noisier, likely due to their fixed mass-to-light ratios and point-source WISE photometry, which underestimates flux. \citetalias{Wen2024ApJS..272...39W}, by contrast, offers wide coverage and high consistency, so we adopt it as our reference. Our results align with recent literature \citep{Kettlety2018MNRAS.473..776K, Palfi2025MNRAS.539.1879P}, which confirms that $W1$-based stellar masses are as robust as SED-based estimates.

In our work, we adopt the \citetalias{Wen2024ApJS..272...39W} methodology, which calibrates stellar mass as a function of $W1$ luminosity and $(r - z)$ color after extinction correction:

$$
\log M_*/M_\odot = \gamma  \cdot \log L_{W1} + a(z_\mathrm{ph}) +  b(z_\mathrm{ph}) \cdot (r_\mathrm{mag} -z_\mathrm{mag}),
$$

where $\gamma$ is a constant and $a(z_\mathrm{ph})$ and $b(z_\mathrm{ph})$ are tabulated by \citetalias{Wen2024ApJS..272...39W}. A similar relation using $z$-band luminosity is also provided. Averaging the two yields a scatter of only 0.15 dex relative to COSMOS2015 \citep{Laigle2016ApJS..224...24L} masses.

Motivated by this, we fit analogous relations between \citetalias{Wen2024ApJS..272...39W} masses and our extinction-corrected photometry for each optical catalog (Pan-STARRS, DELVE, LS DR9, LS DR10). The final relations are detailed in Section \ref{sec:mass_method}.

 \section{Catalog content}
\label{sec:ap2}
Table \ref{tab:column_desc} provides a detailed description of the columns included in REGALADE, while Table \ref{tab:subset} presents its first rows.

 \begin{table}[h!]
  \caption{Column description of the REGALADE catalog. }
 \centering
 \resizebox{\columnwidth}{!}{
 \begin{tabular}{lll}
 \toprule
 Column & Unit & Description \\
 \midrule
 name & &  Identifier from the best distance catalog (ref\_D\_in)\\
 ra & deg &Right Ascension \\
 dec & deg &Declination \\
 id\_cat & & Bitwise flag identifying matched catalogs \\
 R1 & arcsec &Semimajor axis of ellipse \\
 R2 & arcsec &Semiminor axis of ellipse \\
 PA & deg &Position angle of ellipse \\
 ref\_ell & & Index of the catalog used for R1, R2, PA$^{(1)}$\\
 n\_dist & & Number of distance estimates \\
 D & Mpc &Final recommended distance \\
 z & & Final recommended redshift\\
 D\_input & Mpc & Best available distance \\
 D\_err & Mpc & Error on the best available distance  \\
 ref\_D\_in & & Index of the catalog used for D\_input$^{(1)}$ \\
 D\_min & Mpc & Minimum of available distances \\
 D\_max & Mpc & Maximum of available distances \\
 D\_std & Mpc & Standard deviation of available distances\\
 D\_tmean & Mpc & Trimmed mean of available distances\\
 G & mag & Gaia G magnitude \\
 BP & mag & Gaia BP magnitude \\
 PM & mas/yr & Gaia proper motion \\
 angDist & arcsec & Separation to Gaia match \\
 rpsf & mag & r-band PSF magnitude \\
 gmag & mag &$g$-band Kron magnitude \\
 rmag & mag &$r$-band Kron magnitude \\
 imag & mag &$i$-band Kron magnitude \\
 zmag & mag &$z$-band Kron magnitude \\
 W1mag & mag &WISE $W1$ magnitude \\
 W2mag & mag &WISE $W2$ magnitude \\
 griz\_ref & & Catalog used for optical photometry$^{(2)}$ \\
 W\_ref & & Catalog used for infrared photometry$^{(3)}$ \\
 dK & mag & 2MASS extended source K magnitude correction \\
 EBV & mag &Color excess interpolated from \citet{Schlegel1998ApJ...500..525S}\\
 logM & M$_\odot$ & log of the stellar mass\\
 f\_reliability & & Flag galaxies exclusively in lower reliability catalogs \\
 frac\_nearby & & Fraction of annotators that voted "nearby galaxy"\\
 \bottomrule
 \end{tabular}
 }
 \tablefoot{$^{(1)}$ Index as reported in Table \ref{tab:ranking}. $^{(2)}$The griz\_ref column takes values as follows; 0: none, 1: Pan-STARRS, 2: DELVE, 3: LS DR9, 4: LS DR10. $^{(3)}$The W\_ref column takes values as follows; 0: none, 1: AllWISE, 3: LS DR9, 4: LS DR10.}

 \label{tab:column_desc}
 \end{table}

\begin{sidewaystable}

    \caption{First rows of the REGALADE catalog.}
    \centering

    \resizebox{\textwidth}{!}{
\begin{tabular}{cccccccccccccccccccccc}
\hline
name & ra & dec & id\_cat & R1 & R2 & PA & ref\_ell & n\_dist & D & D\_input & ref\_D\_in & D\_min & D\_max & gmag & rmag & zmag & W1mag & griz\_ref & logM & f\_reliability & ... \\\hline
 & deg & deg &  & arcsec & arcsec & deg &  &  & Mpc & Mpc &  & Mpc & Mpc & mag & mag & mag & mag &  & M$_\odot$ &  & ... \\
\hline
DESI J000000.00+285412.8 & 0.0000 & 28.9035 & 9216 & 3.0 & 3.0 & 90.0 & 10 & 2 & 981.4 & 981.4 & 13 & 981.4 & 1419 & 21.6 & 21.0 & 20.7 & 21.2 & 3 & 8.9 & 0& ... \\
WISE J000000.01-151755.9 & 0.0000 & -15.2989 & 2560 & 4.4 & 2.8 & 63.9 & 11 & 2 & 1031 & 1031 & 9 & 1031 & 1699 & 19.2 & 18.1 & 17.4 & 17.2 & 1 & 11.0 & 0& ... \\
DESI J000000.00-524550.5 & 0.0000 & -52.7640 & 50176 & 4.4 & 3.0 & 74.0 & 10 & 3 & 1509 & 1509 & 10 & 1215 & 2179 & 21.3 & 20.6 & 20.3 & 20.5 & 3 & 9.5 & 0& ... \\
DESI J000000.01-485353.8 & 0.0000 & -48.8983 & 50176 & 5.1 & 2.5 & 8.0 & 15 & 3 & 1381 & 1592 & 10 & 935.2 & 1592 & 20.8 & 19.9 & 19.3 & 19.1 & 3 & 10.2 & 0& ... \\
DESI J000000.01-134257.1 & 0.0000 & -13.7159 & 24576 & 3.0 & 3.0 & 0.0 & 13 & 2 & 1703 & 1703 & 13 & 1703 & 2071 & 21.2 & 20.4 & 20.0 & 20.2 & 3 & 9.8 & 0& ... \\
WISE J000000.00-533535 & 0.0000 & -53.5931 & 512 & 3.0 & 3.0 & 0.0 & 9 & 1 & 1101 & 1101 & 9 & 1101 & 1101 &  &  &  & 18.2 & 0 &  & 0& ... \\
WISE J000000.01-472956.4 & 0.0000 & -47.4990 & 50688 & 11.9 & 5.8 & 15.0 & 10 & 4 & 1931 & 1958 & 9 & 1594 & 2113 & 20.2 & 18.6 & 17.7 & 17.0 & 3 & 11.5 & 0& ... \\
DESI J000000.01-451427.7 & 0.0000 & -45.2410 & 50176 & 3.0 & 3.0 & 90.0 & 10 & 3 & 1441 & 1441 & 10 & 791.3 & 2185 & 21.4 & 20.4 & 19.8 & 20.2 & 3 & 9.8 & 0& ... \\
DESI J000000.01-095755.2 & 0.0000 & -9.9653 & 9216 & 4.4 & 3.2 & 70.1 & 10 & 2 & 1984 & 1984 & 13 & 1944 & 1984 & 21.1 & 20.4 & 20.1 & 20.1 & 3 & 9.9 & 0& ... \\
DESI J000000.01-391822.0 & 0.0000 & -39.3061 & 50176 & 7.3 & 4.8 & 163.6 & 10 & 3 & 1416 & 1449 & 10 & 1109 & 1449 & 21.0 & 20.2 & 19.8 & 20.2 & 3 & 9.7 & 0& ... \\
DESI J000000.01-114535.1 & 0.0000 & -11.7598 & 25600 & 3.0 & 3.0 & 90.0 & 10 & 3 & 1897 & 1931 & 13 & 1738 & 1931 & 22.6 & 21.0 & 20.2 & 19.9 & 3 & 10.1 & 0& ... \\
WISE J000000.01+500518.2 & 0.0000 & 50.0884 & 10752 & 4.3 & 3.2 & 201.3 & 13 & 3 & 1136 & 907.7 & 9 & 907.7 & 1136 & 21.0 & 19.8 & 19.0 & 18.6 & 1 & 10.3 & 0& ... \\
DESI J000000.01+130544.9 & 0.0001 & 13.0958 & 46080 & 18.6 & 16.3 & 109.3 & 10 & 3 & 783.5 & 783.5 & 13 & 665.0 & 1422 & 19.9 & 19.3 & 19.0 & 20.1 & 3 & 9.4 & 0& ... \\
DESI J000000.02-724225.1 & 0.0001 & -72.7070 & 49152 & 3.3 & 2.9 & 28.2 & 15 & 2 & 1254 & 1254 & 14 & 1254 & 1552 & 21.5 & 20.9 & 20.5 & 21.1 & 4 & 9.2 & 0& ... \\
WISE J000000.02-330001.2 & 0.0001 & -33.0003 & 50688 & 4.9 & 4.1 & 102.8 & 15 & 4 & 599.9 & 557.4 & 9 & 443.3 & 799.2 & 19.2 & 18.7 & 18.5 & 18.9 & 3 & 9.6 & 0& ... \\
PS J000000.02+450244.8 & 0.0001 & 45.0458 & 10240 & 3.0 & 3.0 & 0.0 & 11 & 2 & 1911 & 1911 & 11 & 1911 & 1911 & 21.5 & 20.8 & 20.2 &  & 1 &  & 1& ... \\
WISE J000000.01+425614.3 & 0.0001 & 42.9373 & 512 & 3.0 & 3.0 & 0.0 & 9 & 1 & 897.9 & 897.9 & 9 & 897.9 & 897.9 & 20.3 & 19.2 & 18.2 & 18.2 & 1 & 10.5 & 0& ... \\
DESI J000000.02-475536.6 & 0.0001 & -47.9268 & 16384 & 3.0 & 3.0 & 0.0 & 14 & 1 & 781.9 & 781.9 & 14 & 781.9 & 781.9 & 22.5 & 21.5 & 20.9 & 21.2 & 3 & 8.9 & 1& ... \\
PS J000000.02-081212.2 & 0.0001 & -8.2034 & 39936 & 6.3 & 2.7 & 45.5 & 10 & 4 & 1465 & 1683 & 11 & 1247 & 1813 & 20.6 & 19.7 & 19.1 & 19.1 & 3 & 10.3 & 0& ... \\
DESI J000000.02+292216.5 & 0.0001 & 29.3713 & 41984 & 4.5 & 3.1 & 26.6 & 10 & 2 & 1795 & 1795 & 13 & 1795 & 1933 & 20.9 & 20.3 & 20.0 & 19.7 & 3 & 9.9 & 0& ... \\
DESI J000000.02+223033.5 & 0.0001 & 22.5093 & 9216 & 3.0 & 3.0 & 90.0 & 10 & 2 & 1673 & 1673 & 13 & 1673 & 1719 & 22.4 & 21.1 & 20.3 & 20.3 & 3 & 9.8 & 0& ... \\
DESI J000000.02-331914.8 & 0.0001 & -33.3208 & 33792 & 3.0 & 3.0 & 90.0 & 10 & 2 & 1728 & 1728 & 10 & 1558 & 1728 & 22.0 & 20.5 & 19.7 & 19.6 & 3 & 10.2 & 0& ... \\
DESI J000000.02-001820.9 & 0.0001 & -0.3058 & 16384 & 3.0 & 3.0 & 0.0 & 14 & 1 & 583.2 & 583.2 & 14 & 583.2 & 583.2 & 24.1 & 23.9 & 23.3 & 19.0 & 4 & 8.5 & 1& ... \\
PS J000000.02+120428.3 & 0.0001 & 12.0745 & 44032 & 4.2 & 2.6 & 147.7 & 10 & 3 & 775.5 & 1586 & 11 & 773.0 & 1586 & 20.4 & 19.8 & 19.3 & 19.7 & 3 & 9.5 & 0& ... \\
PS J000000.02+134354.1 & 0.0001 & 13.7317 & 35840 & 6.7 & 5.0 & 54.0 & 10 & 2 & 798.5 & 798.5 & 11 & 791.7 & 798.5 & 19.6 & 19.1 & 18.7 & 19.3 & 3 & 9.7 & 0& ... \\
...... & ... & ... & ... & ... & ... & ... & ... & ... & ... & ... & ... & ... & ... & ... & ... & ... & ... & ... & ... & ... \\
\hline
\end{tabular}
}
\tablefoot{This table is available in full version at the CDS.}
    \label{tab:subset}
\end{sidewaystable}

\end{document}